\documentclass{JFM-FLM_Au}

\usepackage{graphicx,color}
\graphicspath{{pics}}

\usepackage{epstopdf,epsfig}
\usepackage{newtxtext}
\usepackage{newtxmath}
\usepackage{mathtools}
\usepackage{natbib}

\let\upartial\partial
\newcommand{\pard}[3]{\frac{\upartial^{#3} #1}{\upartial^{#3} #2}}

\usepackage{xparse}

\newcommand{\ep}{\varepsilon}
\newcommand{\ci}{\mathrm{i}}
\newcommand{\dd}{\mathrm{d}}
\newcommand{\ee}{\mathrm{e}}

\providecommand\bcdot{\boldsymbol{\cdot}}
\newcommand{\fb}{\mathbf{f}}
\newcommand{\gb}{\mathbf{g}}

\newcommand{\Gb}{\mathsfbi{G}}

\newcommand{\Ib}{\mathsfbi{I}}
\newcommand{\Jb}{\mathsfbi{J}}

\newcommand{\kb}{\mathbf{k}}

\newcommand{\ub}{\mathbf{u}}

\newcommand{\xb}{\mathbf{x}}

\newcommand\abs[1]{|#1|}
\DeclareMathAlphabet{\pazocal}{OMS}{zplm}{m}{n}

\def\XXint#1#2#3{{\setbox0=\hbox{$#1{#2#3}{\int}$}
     \vcenter{\hbox{$#2#3$}}\kern-.5\wd0}}

\newcommand\Pra{\text{Pr}}

\makeatletter
\def\sgn{\mathop{\operator@font sgn}\nolimits}
\def\Ei{\mathop{\operator@font Ei}\nolimits}
\def\Hi{\mathop{\operator@font Hi}\nolimits}
\makeatother 

\lefttitle{S. Bheemarasetty and S.G. Llewellyn Smith}
\righttitle{On the fundamental solution of viscous internal
  waves. Part 1. Two dimensions}

\title{On the fundamental solution for viscous internal waves and
  Brinkman flows. Part 1. Two dimensions}

\author{Saikumar Bheemarasetty\aff{1}
 \and  Stefan G. Llewellyn Smith\aff{1,2}  \corresp{\email{sgls@ucsd.edu}}}

\affiliation{\aff{1}Department of Mechanical and Aerospace Engineering, Jacobs School of Engineering, University of California San Diego, 9500 Gilman Drive, La Jolla, CA 92093-0411, USA
\aff{2}Scripps Institution of Oceanography, University of California San Diego, 9500 Gilman Drive, La Jolla, CA 92093-0213, USA}

\begin{document}
\maketitle

\begin{abstract}
We obtain the viscous and diffusive fundamental solution for
monochromatic internal waves in a uniformly stratified medium and for
anisotropic Brinkman flow. These solutions take the form of single
integrals with logarithmic singularities, and can be computed
numerically in an efficient manner for possible use in boundary
integral methods. Far-field asymptotic results are
obtained, giving solutions valid far from and inside a ``beam''
corresponding to the internal wave angle in the internal wave case,
consistent with \cite{Thomas:1972}.  For Prandtl numbers $\Pra \gtrsim
O(1)$, the wave field is given by a superposition of wave- and
Stokeslet-like terms. Unlike previous studies, a uniform asymptotic
expansion of the wave-field for $\Pra \gtrsim O(1)$ can be computed
rigorously. Density diffusion attenuates the wave amplitude as to
$(1+\Pra^{-1})^{-2/3}$ and broadens the beam width according to
$(1+\text{Pr}^{-1})^{1/3}$. Evanescent waves in a stratified medium
and anisotropic Brinkman flows have similar behaviour. Anisotropic
Brinkman flow is purely real, dominated by a single circulation
cell. As anisotropy increases, the flow becomes increasingly confined
to the direction with least resistance. The stratified evanescent wave
field has near-vertical cells in its real part, and a dominant single
circulation cell in its imaginary part.
\end{abstract}



\section{Introduction}
\label{sec: intro}

Internal waves are a key feature of geophysical flows. The literature on them is extensive, and we shall not try to review it here beyond a reference to \cite{Sutherland:2010}. A great deal of observational and experimental work has examined their generation, propagation, and interaction with bathymetry, i.e.~scattering. Generation and scattering come from interactions between the stratified fluid and boundary, that is to say fluid-structure interaction, for example in the case of flow over bathymetry \cite[]{Baines:1973,Holloway:1999,SGLS:2002,SGLS:2003, Petrelis:2006,Garrett:2007} or oscillations of objects \cite[]{Mowbray:1967,Gordon:1972,Larsen:1969,Voisin:2011}.

Linear internal waves have an anisotropic dispersion relation. This
has complicated theoretical studies, which have mostly been limited to
simple geometries. To go beyond these to more complicated geometries
requires techniques such as boundary integral methods (BIMs),
themselves the subjects of a vast literature of which we mention only
\cite{Pozrikidis:1992}. \cite{SGLS:2003} solved an integral equation
analytically to find internal tide generation by a steep ridge, but in
general a numerical implementation is needed, as in
\cite{Petrelis:2006}. \cite{Voisin:2021,Voisin:2024b} discuss BIMs for
internal waves. These approaches require a `fundamental solution',
also known as a free-space Green's function, namely a solution of the
flow due to an appropriate isolated singularity.

\cite{Hurley:1972} computed the inviscid Green's function of internal waves in a uniformly stratified non-rotating medium, using analytic continuation. This method solves the hyperbolic governing equation outside the region of internal waves, $\omega > N$ and analytically continues the solution to the internal wave region $\omega<N$. An overview of this method for 2D and 3D internal waves is given in \cite{Martin:2012}, while a detailed investigation of the free-space Green's function are provided in \cite{Voisin:1991}. The inviscid Green's function in a channel is given in \cite{Robinson:1969} and in \cite{Petrelis:2006}.
From \cite{Hurley:1972}, the Green's function for a monochromatic source singularity with frequency $\omega$ located at the origin in a fluid with buoyancy frequency $N$  is
\begin{equation}
	G=-\frac{\ci}{4 \pi N^2 \sin \theta_a \cos \theta_a}\left[\ln \left(x_{+}+\ci 0 \sgn x_{-}\right)+\ln \left(x_{-}+\ci 0 \sgn x_{+}\right)\right], \label{eq:HurleyG}
\end{equation}
where $\theta_a = \cos^{-1}(\omega/N)$, $x_{\pm} = x \cos\theta_a \mp z \sin\theta_a$ and $z_{\pm} = \pm x\sin\theta_a+ z\cos\theta_a$. The phase variations are given by using $\ln(x\pm\ci0) = \ln{|x|} \pm\ci \pi H(-x)$, corresponding to the principal branch of the logarithm.


The wave field (\ref{eq:HurleyG}) has discontinuities between different regions, and BIM methods use a different Green's function that is more singular.  However, experiments going back to \cite{Mowbray:1967} show internal wave beams that spread out without discontinuities far from the source. Using a regularized Green's function can help address these issues, and it is natural to consider viscosity as a regularization method rather than some other \emph{ad hoc} method. A further motivation is that modern experimental techniques, reviewed in e.g.~\cite{Voisin:2011}, need to take into account dissipative effects such as fluid viscosity and the diffusion of heat.

The influence of viscosity was considered by \cite{Thomas:1972}, who used a boundary layer approach to study the wave-field inside an internal wave beam in a viscous stratified medium. This quantified the spread of the beam and obtained a self-similar solution for the wave field inside the beam for small viscosity, but did not give the full wave field throughout the fluid, nor did it include density diffusion. The velocity component parallel to the direction of an internal wave can be written in terms of the similarity solution
\begin{equation}
  u \propto \tilde{x}^{-2/3}\int_0^\infty \zeta \ee^{-\zeta^3} \ee^{-\ci q_0 \zeta} \,\dd\zeta, \qquad q_0 = \tilde{y}\left[\frac{2}{\tilde{x}} \sqrt{\frac{N^2}{\omega^2}-1}\right]^{1/3}, \label{eq:TSG}
\end{equation}
where $\tilde{x}$, $\tilde{y}$ are coordinates nondimensionalized by the length scale $\sqrt{\nu/\omega}$ with $\tilde{x}$ in the direction of group velocity along the internal wave angle $\theta_a = \cos^{-1}(\omega/N)$ measured from the vertical axis, and $\tilde{y}$ is in the opposite direction to the phase velocity.

The effect of viscosity was subsequently considered by \cite{RRao:1977}, \cite{Makarov:1990}, and \cite{Kistovich:2014} among others. It was examined for the generation of the internal tide by \cite{Voisin:2011} and for internal wave excitation by a sphere in \cite{Voisin:2024a}. \cite{Voisin:2003} and \cite{Voisin:2020} include viscosity in the calculation of the free-space Green's function as a perturbative effect. The latter paper in particular provides extensive references to previous work on internal wave generation. These references make approximations in various integrals. \cite{Davis:2010} calculated the wave field generated by a tangentially oscillating disk without explicitly computing the Green's function.

Here  we provide a general integral formulation for the internal wave Green's function in the presence of viscosity and density diffusion and examine the far-field asymptotics, special cases, and the effect of Prandtl number. This approach leads to new, simple expressions, and allows us to obtain numerical and asymptotic results in a rigorous fashion that can be further extended to more complicated problems. The resulting Green's function is the fundamental solution required for BIM implementations.

We also extend our analysis to anisotropic Brinkman flow. Brinkman
flow is an extension to Darcy flow that includes an effective viscosity term. It has applications ranging from biomedical to industrial and geophysical flows. The Green's function for isotropic Brinkman flow can be computed using the Green's function of unsteady Stokes flow \cite[]{Martin:2019}. The situation is more complicated for anisotropic Brinkman flow in which permeability becomes a tensor quantity. \cite{Kohr:2007} provides a formulation for computing this Green's function, but does not show numerical or asymptotic results.

In this paper we consider the two-dimensional monochromatic viscous stratified problem, leaving three-dimensional and other effects for subsequent papers. The general formulation of the problem is given in \S\,\ref{sec: theory}. In \S\,\ref{sec: 2D Gfn}, we obtain tractable expressions for the Green's function and present numerical calculations when $\Pra \to \infty$. The asymptotic behaviour at large distances is presented in \S\,\ref{sec: 2D Asymptotics}, giving expressions valid away from and near to the wave angle, as well as a rationally obtained uniform asymptotic expansion valid for all observer angles. The evanescent case and special limits are presented in \S\,\ref{sec: special cases}. In \S\,\ref{sec: 2D Gfn Pr} we consider the effect of Prandtl number. Results for anisotropic Brinkman flow are presented in \S \ref{sec: brinkman}, while \S\,\ref{sec:conc} concludes. Some technical results are given in the Appendices.

\section{Green's function formulation}
\label{sec: theory}

We consider two-dimensional monochromatic problems with a factor of $\ee^{-\ci\omega t}$ (taking $\omega$ real and positive without loss of generality) in all variables. The viscous governing equations in the Boussinesq approximation for stratified flow and no rotation are
\begin{eqnarray}
-\ci\omega u_i & = & -\frac{1}{\rho_0} \pard{p}{x_i}{} + \nu \nabla^2
u_i + b \delta_{i3} +  g_i \delta(\xb - \xb_0), \label{u eqn pole}\\
-\ci\omega b + N^2 u_3 & = & {D} \nabla^2 b, \label{b eqn pole}\\
\pard{u_i}{x_i}{} & = & 0,
\label{incomp pole}
\end{eqnarray}
in suffix notation using the Einstein summation convention, with $u_i$ the  components of velocity ($i = 1$ and $3$), $p$ pressure, $b$ buoyancy, $\nu$ kinematic viscosity, $\rho_0$ a reference density of the fluid, $D$ buoyancy diffusivity and $N$ buoyancy frequency. We take $N$ to be constant so that the stratification is uniform. Following \cite{Pozrikidis:1992}, we have introduced a singularity in the momentum equations (\ref{u eqn pole}), which gives the appropriate Green's function for use in BIMs. We write the associated velocity field as
\begin{equation}
u_i = \frac{1}{4\pi\nu}G_{ij}(\xb,\xb_0) g_j.
\end{equation}
The dimensions of $g_i$ are length$^3$/time$^2$, while the elements $G_{ij}$ of $\Gb$ are dimensionless.

We define the Fourier transform of $f(\xb)$ by
\begin{equation}
(\mathcal{F} f)(\kb) = \int f(\xb) \ee^{-\ci\kb\cdot\xb} \,\dd\xb. \label{Fourier transform}
\end{equation}
Then, with $U_i$, $P$, $B$ the Fourier transforms of $u_i$, $p$ and $b$ respectively, we have
\begin{equation}
-\ci\omega U_i = -\frac{\ci k_i}{\rho_0} P - \nu |\kb|^2 U_i + B
\delta_{i3} +  g_i \ee^{-\ci\kb\cdot\xb_0}, \qquad -\ci\omega B + N^2
U_3 = -{D}|\kb|^2 B, \qquad k_i U_i = 0. \label{Fourier eqns}
\end{equation}
Now eliminate $P$ and $B$ to obtain equations for the $U_i$. These can be combined to give the Fourier transform of $\Gb(\xb,\xb_0)$ as
\begin{equation}\everymath{\displaystyle}
(\mathcal{F} \Gb)(\kb) = 4\pi\frac{\nu}{\omega} \frac{\gamma}{\alpha} \left(\begin{array}{ll}
k_z^2 & -k_xk_z \\
-k_xk_z & k_x^2
\end{array} 
\right) \ee^{-\ci \kb\bcdot\xb_0},   \label{2D Greens fn with pr}
\end{equation}
where the quantities $\alpha$ and $\gamma$ are given by
\begin{equation}
\gamma = 1 + \frac{\ci D}{\omega} |\kb|^2, \qquad  \omega \alpha = \gamma\left(\nu|\kb|^4 - \ci\omega|\kb|^2 \right) + \ci\frac{N^2}{\omega} |\kb_H|^2. \label{abgdim}
\end{equation}
 The symmetry property $G_{ij} =  G_{ji}$ is apparent. This can also be shown from the original equations and is a form of the Lorentz reciprocity relation.

Applying the inverse Fourier transform gives
\begin{equation}
    \Gb = \frac{1}{\pi} \int_{-\infty}^\infty \int_{-\infty}^\infty
    \frac{\gamma}{\alpha} \begin{pmatrix}
        m^2 & -km\\
        -km &k^2
    \end{pmatrix} \ee^{\ci \lambda\kb\bcdot\hat{\xb}} \,\dd k \,\dd m,
    \label{eq:Gb1}
  \end{equation}
  where now the quantities (\ref{abgdim}) are replaced by  the dimensionless forms
\begin{equation}
\gamma = 1 + \ci \Pra^{-1} |\kb|^2, \qquad \alpha = \gamma\left(|\kb|^4 - \ci |\kb|^2\right) + \ci\frac{N^2}{\omega^2} |\kb_H|^2 \label{abg}
\end{equation}
and $\kb = (k,m)$.  In the exponential we write $R = |\xb-\xb_0|$
for the dimensional distance from the source to the observer, and
$\hat{\xb} = (\xb - \xb_0)/|\xb - \xb_0|$ for the unit vector from the
source to the observer. Then  $\lambda = R \sqrt{\omega/\nu}$ is the
usual dimensionless distance, recovering the near-singularity Stokeslet
results.

This dispersion relation $\alpha = 0$ is a cubic in $k^2$ and $m^2$. As $\Pra \to \infty$, one recovers the dispersion relation of viscous internal waves, as in \cite{Davis:2010}. The inviscid dispersion relation comes from taking $\nu = D = 0$, and is given by $\omega = N\cos{\theta}$,	where $\theta$ is the angle between the group velocity of internal waves (the energy propagation direction) and the vertical. In the rest of this section and \S\,\ref{sec: 2D
		Gfn}--\ref{sec: special cases} we take  Pr$\to \infty$.

\begin{figure}
  \centering
  \includegraphics[width=\textwidth]{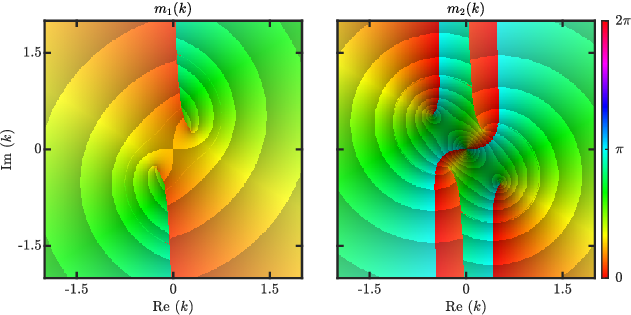}
  \caption{Phase portraits of the roots of $\alpha$ in the $k$-plane for $\omega/N = 0.8$. The colour bar corresponds to the phase angle.}
  \label{fig: alpha roots}
\end{figure}

We can perform the $m$-integral using the calculus of residues. The
singularities in the integrand are at zeros of $\alpha$, with
\begin{equation}
\alpha = (k^2+m^2)^2 -\ci(k^2 + m^2) + \ci
\frac{N^2}{\omega^2} k^2. \label{alpha beta in m,k}
\end{equation}
Viewed as an equation in $m$, this is a biquadratic with four roots in the complex plane:
\begin{equation}
    m_1 = \ci\sqrt{k^2 -\frac{\ci + \ci\sqrt{1+4\ci k^2(N/\omega)^2}}{2}}, \qquad m_2 = \ci\sqrt{k^2 - \frac{\ci - \ci\sqrt{1+4\ci k^2(N/\omega)^2}}{2}}, \label{m alpha eqn}
\end{equation}
along with $-m_1$ and $-m_2$. We pick the branch of square root in the
complex plane with positive real part so as to define $m_j$ as a single-valued function with complex argument $k$. Then $m_j$ is the branch with positive imaginary part for all $k$. The branch cut structure of the roots of $\alpha$ is given in Figure~\ref{fig: alpha roots}

 The difficulty of computing the Green's function lies in taking the inverse Fourier transform. After calculating the integral in $m$, we are left with an integral in $k$. This is a bad idea because the zeros of $\alpha$ result in a double square root and the $k$-integral takes the form
\begin{equation}
    I(\hat{\xb};\lambda) = \int_C f(k) \ee^{\lambda h(k;\hat{\xb})} \,\dd k, \qquad   h(k; \hat{\xb}) = \ci k \hat{x} +  \ci m_{1,2}(k) |\hat{z}|, \label{Model Integral}
\end{equation}
where $C$ is a contour in the $k$-plane.  For small viscosity and in the far-field, $\lambda \gg 1$, leading to highly oscillatory integrals. The modulus sign in $|\hat{z}|$ is needed to guarantee convergence of the $k$-integrals. This integral is analytically and numerically difficult. Instead of dealing with it and its complicated branch-cut structure, we use polar coordinates.

\section{Green's function with $\Pra \to \infty$ }
 \label{sec: 2D Gfn}
\subsection{Formulation}
The Green's function (\ref{eq:Gb1}) becomes
\begin{equation}
    \Gb = \frac{1}{\pi} \int_{-\infty}^\infty \int_{-\infty}^\infty \begin{pmatrix}
        m^2 & -km\\
        -km &k^2
    \end{pmatrix} \frac{\ee^{\ci \lambda\kb\bcdot\hat{\xb}} }{(k^2+m^2)^2 -\ci(k^2 + m^2) + \ci (N^2/\omega^2) k^2} \,\dd k \,\dd m.
\end{equation}
The diagonal elements of $G_{ij}$ are symmetric under reflection across the $x$-axis and $z$-axis. All of its elements are symmetric under rotation by $\pi$ about the origin in the $(x,z)$ plane. We make the unconventional definition $(\hat{x},\hat{z}) = (-\sin{{\theta}_d}, \cos{{\theta}_d})$, so that $\theta_d$ denotes the angle measured in the positive sense from the vertical axis to $\hat{\xb}$. Because of the symmetries we can limit $\theta_d$ to the range $(0,\pi/2)$ and construct the solution elsewhere using symmetry.

Using polar coordinates $(k,m) = \kappa (\cos{\theta},\sin{\theta}) \equiv \kappa(c,s)$, we find
\begin{align}
   \Gb  &= \frac{1}{\pi} \oint \begin{pmatrix}
        s^2 & -sc\\
        -sc &c^2
    \end{pmatrix} \,\dd\theta \int_0^\infty  \frac{\kappa }{\kappa^2 +\ci\left({N^2c^2}/{\omega^2} -1\right)}\ee^{\ci \lambda\kappa(c\hat{x} + s\hat{z})} \,\dd\kappa \\
    &= \oint  \fb(\theta) K(a(\theta),d(\theta)) \,\dd\theta, \label{theta integral in I}
\end{align}
where the $\kappa$-integral $K(a(\theta),d(\theta))$ is defined as
\begin{equation}
     K(a(\theta),d(\theta)) = \int_0^\infty \frac{\kappa}{\kappa^2 + a^2(\theta)} \ee^{\ci \lambda d(\theta)\kappa} \,\dd\kappa \label{kappa integral IW}
\end{equation}
and the components of $\fb$ are
\begin{equation}
  \frac{1}{\pi} \left( \begin{array}{cc}
       \sin^2{\theta} & -\sin{\theta} \cos{\theta} \\
      -\sin{\theta} \cos{\theta} & \cos^2{\theta}
                       \end{array}
                     \right).
\end{equation}

Let $\theta_a = \cos^{-1}{(\omega/N)}$, corresponding to the angle made by an inviscid plane internal wave. Without loss of generality we have $0 \leq \theta_a \leq \pi/2$. The steady and critical limits $\theta_a \to \pi/2$ and $\theta_a \to 0$ are considered in \S\,\ref{sec:steady} and \S\,\ref{sec:critical}.  Then
\begin{align}
a^2(\theta) &= \ci\left(\frac{N^2}{\omega^2}{\cos^2{\theta}} -1\right) = -\frac{\ci}{\cos^2{\theta_a}}\sin{(\theta-\theta_a)}\sin{(\theta+\theta_a)}, \label{eq:adef}\\
d(\theta) &= \hat{x} \cos{\theta} + \hat{z}\sin{\theta}=  \sin{(\theta - {\theta_d})}. \label{eq:ddef}
\end{align}
The real function $d(\theta)$ has period $2\pi$, while $a^2(\theta)$
is purely imaginary with period $\pi$. The function $a(\theta)$ is multivalued, and we take the branch with $\Real\;{a} >0$.  The integral $K(a(\theta),d(\theta)) $ does not exist when $a(\theta) = 0$ or $d(\theta) = 0$, corresponding to $\theta \in \Theta_a$ and $\theta \in \Theta_d$, respectively, with
\begin{equation}
    \Theta_a = \{-\pi +\theta_a, -\theta_a, \theta_a, \pi - \theta_a\}, \qquad \Theta_d =  \{\theta_d,\pi+\theta_d\}.
    \label{a0, b0 eqns}
\end{equation}
With detailed algebra relegated to Appendix~\ref{appendix: 2D IW kappa integrals}, the $\kappa$-integrals yield, with $K(a,d) = I(ad) + J(a,d)$,
\begin{align}
  I(ad) &= \frac{1}{2} [\ee^{\lambda ad}E_1(\lambda ad) +  \ee^{-\lambda ad}E_1(-\lambda ad)], \label{eq:I}\\
  J(a,d) &= \begin{cases} 0 &\mbox{if $\sgn{d}\arg a \in (0,\pi/2)$} \\
\sgn{d}(\ci\pi)\ee^{-\lambda a|d|} & \mbox{if $\sgn{d}\arg a \in (-\pi/2,0)$}, \label{eq:J}
\end{cases} 
\end{align}
with $E_1(z)$ the exponential integral function \cite[]{AS}. The function $E_1(z)$ has a branch cut along the negative real axis
and a logarithmic singularity at the origin, which here corresponds to
$\lambda ad = 0$, that is the values of $\theta$ in $\Theta_a$ or $\Theta_d$. The function $I(ad)$ is even in $ad$. We define $\Ib$ and $\Jb$ as the $\theta$-integrals with weight $\fb$ of $I(ad)$ and $J(a,d)$.

\begin{figure}
  \centering
  \includegraphics[width=0.6\linewidth]{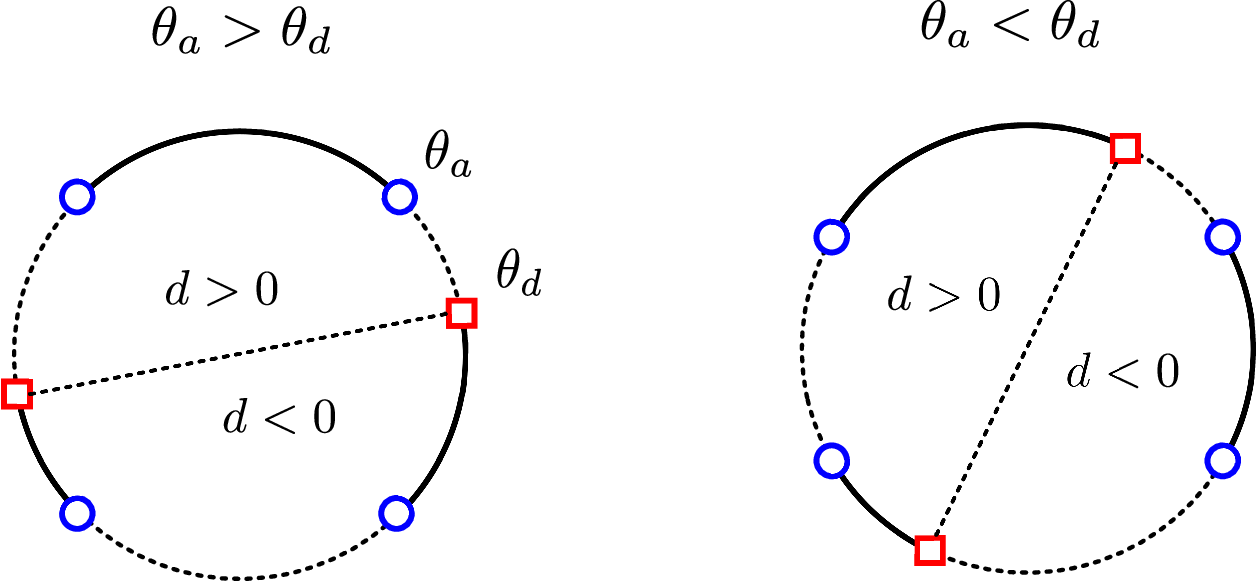}
  \caption{Singularities in the $\theta$-integral for two different observer locations. The circles corresponds to $\Theta_a$ and the rectangles to $\Theta_d$. The solid curves correspond to nonzero $J$, i.e.~$\sgn{d}\arg a \in (-\pi/2,0)$, and the dotted curves to $J = 0$.}
  \label{fig: theta angles}
\end{figure}

The singularities $\Theta_a$ and $\Theta_d$ on the unit circle are shown in Figure~\ref{fig: theta angles}, along with the intervals along which $J(a,d)$ does not vanish. If the observer is exactly at the internal wave angle, so that $\theta_d = \theta_a$, we have $\Theta_d \subset \Theta_a$.

\subsection{Numerical results}
\label{subsec: Numerical limitations}
The integral (\ref{theta integral in I}) can be computed in a number of ways. The integrand has logarithmic singularities, and removing
them makes the calculation better behaved, but is not needed for
modest precision. The singularities can be removed using
the results
\begin{equation}
  \oint \log{\left|{\frac{N^2}{\omega^2} \cos^2{\theta} -1}\right|}
  \,\dd\theta  = 4\pi\log{\left(\frac{N}{2\omega}\right)}, \qquad
  \oint \sin{\theta}\cos{\theta} \log{\left|\frac{N^2}{\omega^2} \cos^2{\theta} -1\right|}\,\dd\theta  = 0, \label{log identity 1}
\end{equation}
and
\begin{equation}
  \oint \log{|\sin{(\theta - {{\theta}}_d)}|} \,\dd\theta  = -2\pi\log{2}. \label{log identity 2}
\end{equation}
 To evaluate the $\theta$-integrals, the interval was divided into subintervals between the zeros of $a(\theta)$ and $d(\theta)$, and in each interval, a Gauss--Legendre quadrature rule with $250$ nodes was applied. Using the \texttt{integral} function in {\sc Matlab} gave results that agreed to 10 decimal places.

\begin{figure}
  \centering
  \includegraphics[width=\textwidth]{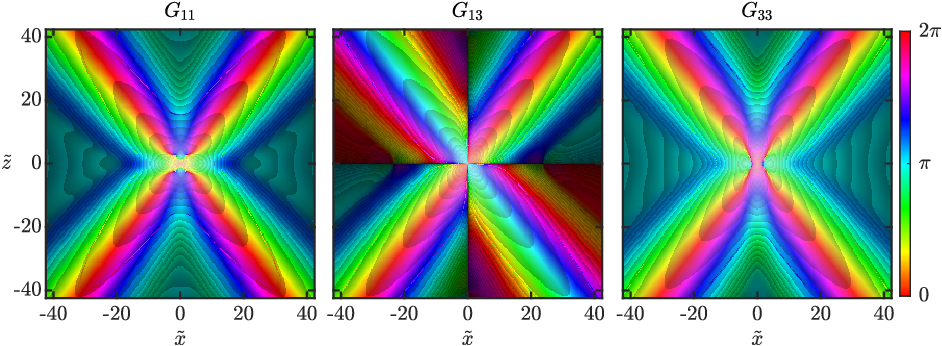}
  \caption{Phase portrait of $\Gb$ for $\omega/N = 0.8$ plotted against the dimensionless displacement from the origin $\tilde{\xb} = (\xb - \xb_0) \sqrt{\omega/\nu}$.}
  \label{fig:2D Greens function kappa integrals}
\end{figure}

The complex phase portrait of the Green's function is shown in Figure~\ref{fig:2D Greens function kappa integrals}, for $\omega/N = 0.8$, i.e.~$\theta_a \approx 36.87^\circ$. The contour lines correspond to constant magnitude. Notice that $G_{1,2}$ is asymmetric about the $x$- and $y$-axes, which can be seen as the phase portrait gives black lines along the axes. while $G_{11}$ and $G_{22}$ are symmetric. The amplitude of the components decays, although this is hard to see. This will be quantified later.

\section{2D Asymptotic behaviour: $\lambda \to \infty$}
\label{sec: 2D Asymptotics}

\subsection{Approach}
\label{subsec:approach}
Studying the asymptotic behaviour of the 2D Green's function as $\lambda\to\infty$ requires different approaches for $I$ and $J$ in (\ref{eq:I}--\ref{eq:J}) in a formal sense, although the approaches are actually quite similar. For $I$ we look at the behaviour of the argument $\lambda a(\theta) d(\theta)$. This argument is large except near the zeros of $a(\theta)$ or $d(\theta)$, so that away from these zeros we can take the limiting behaviour of $I$ to evaluate the integral. We hence split the range of integration in $\theta$ into a sum of local contributions, corresponding to the regions near singularities, and the remaining global contribution, corresponding to the rest of the range of integration. This approach, sometimes called ``divide and conquer,'' is explained most clearly in \cite{Hinch:1991}. The local contributions are taken to come from intervals of width $2\delta_i$ about each singularity $\tilde{\theta}_i$, where the $\delta_i$ are parameters that have to be sufficiently small, but that cannot affect the final result. Then the $\theta$-integral of $I$ can be written as 
\begin{equation}
  \oint \fb(\theta) I(a(\theta)d(\theta)) \,\dd\theta = \left[ \int_G + \sum_{i=1}^6 \int_{\tilde{\theta}_i -\delta_i}^{\tilde{\theta}_i + \delta_i} \right] \fb(\theta) I(a(\theta)d(\theta)) \,\dd\theta. \label{splitting integral}
\end{equation}
                               
The individual local and global integrals are divergent, but since the original integral exists, the divergent terms from the local and global contributions will cancel. Rather than tracking these cancellations, we can use Hadamard finite-part integrals \cite[see e.g.][]{Wong:2001,Estrada:2012} to take care of this automatically, something that does not seem to be discussed in standard textbooks. From now on, any diverging integrals will be taken to be Hadamard finite parts, and we drop the abbreviation $\mbox{F.p.}$ for convenience. Hadamard finite-part integrals can be computed using the result
\begin{equation}
  \int_C F(z) \,\dd z = 2\pi\ci\sum_{j=1}^n \mbox{Res}_{z=b_j}
  F(z) + \pi\ci\sum_{k=1}^m \mbox{Res}_{z=a_k} F(z), \label{Hadamard residues}
\end{equation}  
for a closed contour $C$, where the $a_1,\ldots,a_m$ are poles of the
function $F$ on the contour and $b_1,\ldots,b_n$ are isolated
singularities inside the contour. This formula does not hold if there
are essential singularities on the contour.

The asymptotics of $J$ can be obtained from Laplace's method. This is also a local method, since the dominant contributions come from the zeros of $\lambda a(\theta) d(\theta)$ again, but since the global contributions are exponentially small they can be ignored. We now consider $I$ and $J$ separately, writing $\ep = \lambda^{-1}$. 

\subsection{Global contribution to $\Ib$}
\label{subsec: global contribution}
From (\ref{eq:I}) and the asymptotic expansion of $E_1$,
\begin{equation}
     \ee^z E_1(z) \sim \frac{1}{z} \left( 1 - \frac{1!}{z} + \frac{2!}{z^2} - \frac{3!}{z^3} + \dots\right),  \qquad z \to \infty, \quad |\arg z| < \frac{3\pi}{2}, \label{e^z E_(z) asym}
\end{equation}
the behaviour of $I$ for large $\lambda$ is found to be
\begin{equation}
\Ib_{g} \sim -\ep^2\oint
\frac{\fb(\theta)}{a^2(\theta)d^2(\theta)}\,\dd\theta-\ep^4\oint
\frac{3! \fb(\theta)}{a^4(\theta)d^4(\theta)}\,\dd\theta + \cdots = O(\ep^\infty), \label{Global contribution series}
\end{equation}
meaning that $\Ib_{g}$ is zero to all orders in $\ep$. Closing the contours in
(\ref{Global contribution series}) in the upper or lower half-planes
shows that all the integrals vanish on using (\ref{Hadamard residues}). The forms (\ref{e^z E_(z) asym}) and (\ref{Global contribution
  series}) are not valid close to the zeros of $a$ and $d$, so a separate local
calculation is also required, which will be given in \S\,\ref{subsec:
  local contribution off}.

\subsection{Contribution to $\Jb$} 
The integral $\Jb$ can be treated using Laplace's method. Its behaviour
for large $\lambda$ is dominated by regions near the zeros of the
argument of the exponential, which has period $\pi$. Close to $\theta_a$, we write $\theta = \theta_a + \ep^2 \tau$, and find from \eqref{eq:adef}
\begin{equation}
    \lambda a({\theta}) d({\theta}) = \sqrt{\xi_1(\theta_a) \tau}
    \sin{(\theta_a - \theta_d)} + O(\ep^2) , \label{lad theta_a exp1}
  \end{equation}
where $\xi_1(\theta_a) = -2\ci\tan{\theta_a}$, while close to $-\theta_a$,
\begin{equation}
    \lambda a(\theta)d(\theta) = -\sqrt{\xi_1(-\theta_a) \tau}
    \sin{(\theta_a + \theta_d)} + O(\ep^2). \label{lad theta_a exp2}
  \end{equation}
Laplace's method hence yields, for the contribution from the pairs of
singularities $(\theta_a,\pi+\theta_a)$, $(-\theta_a,\pi - \theta_a)$,
\begin{equation}
    \Jb_{\{a,-a\}} = -\ep^2 2\pi \cot{\theta_a}
    \left[\frac{\fb(\theta_a)}{\sin^2(\theta_a-\theta_d)} +
      \frac{\fb(-\theta_a)}{\sin^2(\theta_a+\theta_d)}\right]
    + O(\ep^4). \label{eq:Jaa}
\end{equation}
Using the expansion
\begin{equation}
    \lambda a({\theta}) d({\theta}) = \pm [\lambda a(\theta_d) \ep
      \tau + \lambda a'(\theta_d) \ep^2\tau^2  +
      \cdots], \label{lad theta_d exp}
  \end{equation}
with the plus sign for $\theta_d$ and the minus sign for $\pi+\theta_d$,
the contribution from $(\theta_d,\pi+\theta_d)$ is
\begin{equation}
    \Jb_{d} = \mp \ep 2\pi\ci \frac{\fb(\theta_d)}{a(\theta_d)} +
    0\ep^2+ {O}(\ep^3), \label{eq:Jd}
\end{equation}
where the plus sign corresponds to $\theta_a > \theta_d$ and the minus
sign to $\theta_a<\theta_d$ (see figure~\ref{fig: theta angles}).

This result can be viewed as local, since it considers contributions
from the neighbourhoods of the zeros of $\lambda ad$, but the contributions
from the rest of the integration range are exponentially small. Hence
a global calculation is not needed. These asymptotic results fail when
$\theta_d$, the angle to the observer, approaches $\theta_a$, the
internal wave angle, so the resulting asymptotic expansion is not
uniform in that region, which we call the ``beam.'' The region close to the beam will
be considered in \S\,\ref{subsec: local contribution}.

\subsection{Local contribution to $\Ib$ off the beam}
\label{subsec: local contribution off}
We consider contributions from the three pairs of singularities at
$(\theta_a,\pi+\theta_a)$, $(-\theta_a,\pi - \theta_a)$,
$(\theta_d,\pi+\theta_d)$ separately. From (\ref{lad theta_a exp1})
and the symmetry of $I(ad)$, the local contributions from $\theta_a$
and $\theta_a + \pi$ together give
\begin{equation}
\Ib_{l,a} = \ep^2 {\fb(\theta_a)}\int_{-\delta/\ep^2}^{\delta/\ep^2} [\ee^{h(\tau)} E_1(h(\tau)) + \ee^{-h(\tau)} E_1(-h(\tau))] \,\dd\tau + O(\ep^4),
\end{equation}
where
$h(\tau) = \sin(\theta_a-\theta_d) \sqrt{\xi_1(\theta_a) \tau}$ and
$\ep^2 \ll \delta$. This is a principal-value integral with an
integrable singularity at the origin that can be computed exactly
using the result
\begin{equation}
    \int_{-M}^M \left[\ee^{cx^{1/2}} E_1(cx^{1/2}) +\ee^{-cx^{1/2}} E_1(-cx^{1/2})\right] \,\dd x =  \frac{1}{c^2}\begin{cases}
        2\pi\ci &\mbox{for $\arg{c} \in (0,\pi/2)$}, \\
        -2\pi\ci &\mbox{for $\arg{c} \in (-\pi/2, 0)$}.
    \end{cases}
\end{equation}
Hence
\begin{equation}
\Ib_{l,a}= \ep^2 \pi \frac{\fb(\theta_a) \cot{\theta_a}}{
  \sin^2{(\theta_a - \theta_d)}} + O(\ep^4). \label{eq:Ila}
\end{equation}

The local contribution from the zeros at $\{-\theta_a,
\pi-\theta_a\}$ is obtained similarly. The argument $\lambda
a(\theta)d(\theta)$ close to the zero at $-\theta_a$ is given
by (\ref{lad theta_a exp2}). Proceeding as above gives
\begin{equation}
    \Ib_{l,-a} = \ep^2 \pi \frac{\fb(-\theta_a) \cot{\theta_a}}{
  \sin^2{(\theta_a + \theta_d)}} + O(\ep^4). \label{eq:Il-a}
\end{equation}

The third contribution comes from $\{\theta_d,\pi + \theta_d\}$. Using
(\ref{lad theta_d exp}) we find
\begin{align}
    \Ib_{l,d} =& \ep {\fb(\theta_d)}\int_{-\delta/\ep}^{\delta/\ep} [\ee^{h(\tau)} E_1(h(\tau)) + \ee^{-h(\tau)} E_1(-h(\tau))] \,\dd\tau + 0 \ep^2 + O(\ep^3), \label{eq:Ibld}
\end{align}
with $h(\tau) = a(\theta_d) \tau$, as the integrals at $O(\ep^2)$ are
odd functions of $\tau$ and hence vanish by symmetry. The $O(\ep)$
integral can be computed using integration by parts, giving
\begin{equation}
    \Ib_{1,d} = \pm \ep 2\pi\ci \frac{\fb(\theta_d)}{a(\theta_d)} +
    0\ep^2 + {O}(\ep^3), \label{eq:Ild}
\end{equation}
where the positive sign corresponds to $\theta_a > \theta_d$ and the negative sign to $\theta_a<\theta_d$. Note that $\Ib_{1,d} + \Jb_{1,d} = O(\ep^3)$, i.e.~the combined contribution from $\theta_d$ and $\pi + \theta_d$ cancels at $O(\ep)$ and $O(\ep^2)$.

Combining (\ref{eq:Jaa}), (\ref{eq:Jd}), (\ref{eq:Ila}),
(\ref{eq:Il-a}) and (\ref{eq:Ild}) gives the leading-order velocity-field
\begin{equation}
    \Gb \sim \Ib_l \sim -\ep^2\pi \left[ \frac{\fb(\theta_a) \cot{\theta_a}}{\sin^2{(\theta_a - \theta_d)}} + \frac{\fb(-\theta_a) \cot{\theta_a}}{ \sin^2{(\theta_a + \theta_d)}} \right]. \label{eq:offbeamas}
\end{equation}
The $O(\ep)$ terms have cancelled. The components of $\Gb$ are purely real at leading order. This result can be rewritten in terms of the original variables as
\begin{equation}
  \Gb \sim -\mu^2 \left[ \left(
      \begin{array}{rr}
        1 - \varpi^2 & -\varpi \sqrt{1-\varpi^2} \\
        -\varpi \sqrt{1-\varpi^2} & \varpi^2
      \end{array} \right) \frac{1}{\sigma_-^2} + \left(
      \begin{array}{rr}
        1 - \varpi^2 & \varpi \sqrt{1-\varpi^2} \\
        \varpi \sqrt{1-\varpi^2} & \varpi^2
      \end{array} \right) \frac{1}{\sigma_+^2} \right], \label{eq:GBasorigv}
\end{equation}
writing $\varpi = \omega/N$ for brevity and using the modified coordinates
\begin{equation}
  \sigma_\pm = \hat{x} \varpi   \mp \hat{z} \sqrt{1 - \varpi^2} =  \hat{x} \cos{\theta_a} \mp \hat{z}\sin{\theta_a} 
\end{equation}
similar to those \cite{Hurley:1972}. The variable
\begin{equation}
      \mu = \left(\frac{\ep^2 \varpi}{\sqrt{1 - \varpi^2}}\right)^{1/2} = \frac{(\nu/N)^{1/2}}{R (1 - \varpi^2)^{1/4}}
\end{equation}
turns out to be a more convenient variable to understand the physical behaviour of the solution. The expression (\ref{eq:GBasorigv}) fails as $\sigma_\pm$ approach zero, corresponding to the near-beam case. In the steady limit $\varpi \to 0$, $\mu$ and the entries in the matrix are finite; see \S\,\ref{sec:steady}. For the case $\varpi \to 1$, the behaviour will depend on $\mu$; see \S\,\ref{sec:critical}.

\subsection{On-beam asymptotics}
\label{subsec: local contribution} 
Near the beam we write  $\theta_a -\theta_d = q \ep^{\eta}$ with $\eta > 0$. Rather than (\ref{lad theta_a exp1}), the argument $\lambda a(\theta) d(\theta)$ then becomes
\begin{equation}
   \lambda a({\theta}) d({\theta}) 
   =\lambda \sqrt{\frac{\xi_1(\theta_a)}{\sin{(2\theta_a)}}} \sqrt{\sin{(\sigma \tau)}\sin{(2\theta_a + \sigma \tau)}} \sin{(q \ep^\eta + \sigma \tau)}. \label{lam ad original}
 \end{equation}
Expanding about $\theta_a$ using $\theta = \theta_a + \sigma \tau$ gives
\begin{equation}
    \lambda a({\theta}) d({\theta}) =\sqrt{\xi_1(\theta_a)}[\lambda \sigma^{3/2} \tau^{3/2} + q\lambda^{1-\eta} \sqrt{\sigma} \sqrt{\tau} ] + \cdots. \label{lam ad sig eta}
\end{equation}
The dominant balance between these two terms gives $\eta = 2/3$ and $\sigma = \ep^{2/3}$ so that both terms contribute at $O(1)$,

Hence we obtain a wave-field of $O(\ep^{2/3})$ near the internal wave beams, which we define by $|\theta_a -\theta_d| \lesssim O(\ep^{2/3})$. We can quantify the beam spread by $R |\theta_a - \theta_d| \propto R\ep^{2/3}  \propto\left({R\nu}/{\omega}\right)^{1/3}$, since $\lambda = R\sqrt{\omega/\nu}$. This beam width is consistent with \cite{Thomas:1972}.

\begin{figure}
  \centering
  \includegraphics[width=0.8\linewidth]{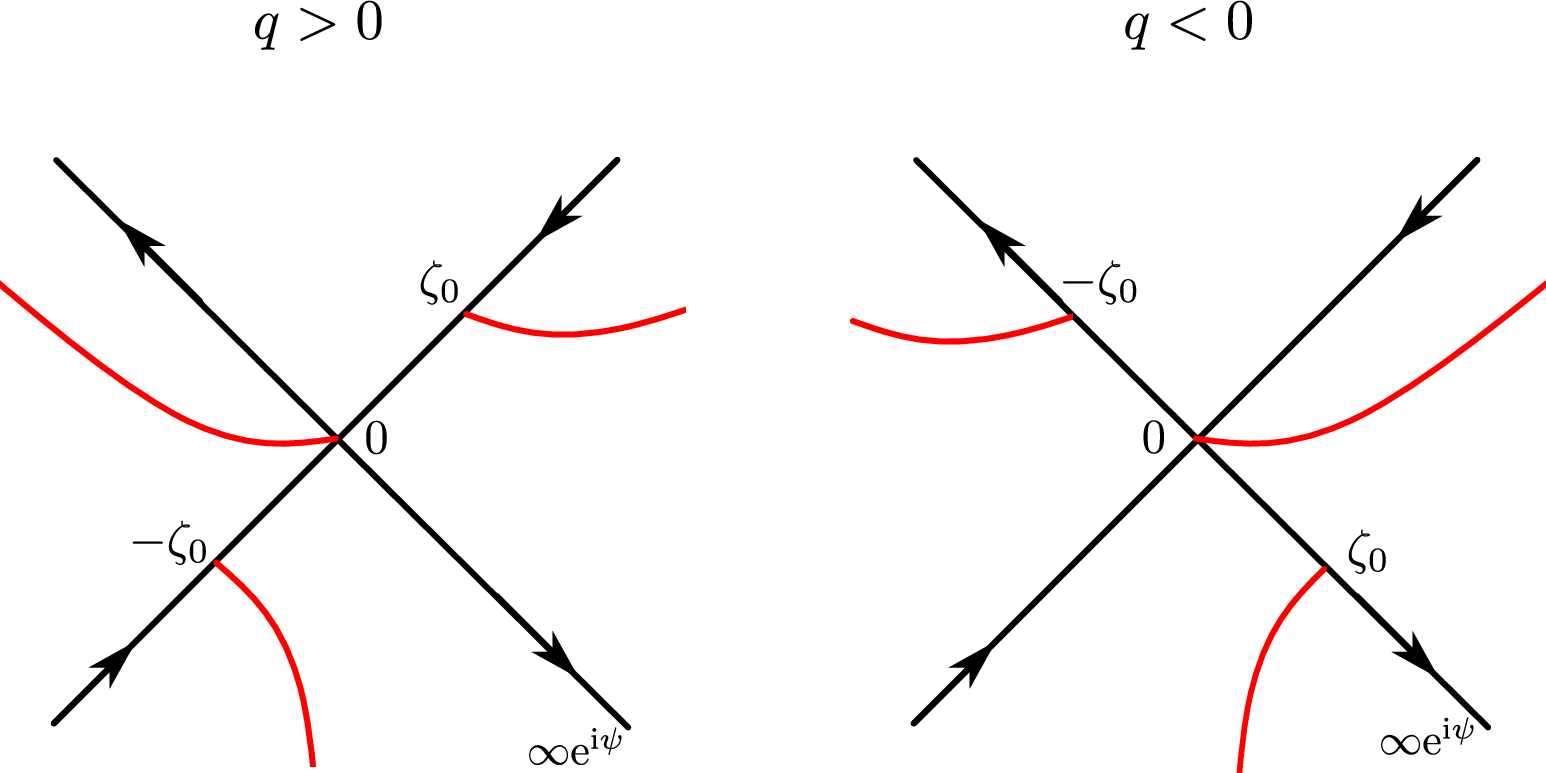}
  \caption{Schematic of integration contours (black lines) in the
    $\zeta$-plane and branch cut structure (red curves) of $\log{(q
    \zeta + \xi_1^{-1} \zeta^3)}$ for $q \lessgtr 0$.}
  \label{fig: Contours}
\end{figure}

The global contribution remains exponentially small, since the analysis of \S~\ref{subsec: global contribution} holds for all observer angles. Near the beam we use the expansion
\begin{equation}
  \lambda a(\theta)d(\theta)  = \sqrt{\xi_1(\theta_a)} \sqrt{\tau} (q+\tau) + {O}(\ep^{2/3}), \label{lam adin q}
\end{equation}
valid for near the singularities $(\theta_a,\pi+\theta_a)$.  Using ($\ref{lam adin q}$), the local contribution to $I$ is given by
\begin{align}
\Ib_l = &\left[\int_{{\theta}_a -\delta}^{{\theta}_a + \delta} + \int_{\pi+{\theta}_a -\delta}^{\pi + {\theta}_a + \delta} \right] \fb(\theta) K(a(\theta),d(\theta)) \,\dd\theta \\
\     &= \ep^{2/3} \fb(\theta_a) \int_{-\delta/\ep^{2/3}}^{\delta/\ep^{2/3}} [\ee^{h(\tau)}E_1(h(\tau)) + \ee^{-h(\tau)}E_1(-h(\tau))] \,\dd\tau + O(\ep^{4/3}),
\end{align}
where $h(\tau) = \sqrt{\xi_1(\theta_a)} \sqrt{\tau}(q+\tau)$ and $\ep^{2/3} \ll \delta$. The change of variables $\zeta = \sqrt{\xi_1\tau}$ leads to
\begin{equation}
\Ib_l = \ep^{2/3} \frac{2 \fb(\theta_a)}{\xi_1(\theta_a)} \left[ \int_{{M} \ci\ee^{\ci \psi}}^{{M} \ee^{\ci \psi}} + \int_{-{M}\ci\ee^{\ci \psi}}^{- {M}\ee^{\ci \psi}}   \right] \zeta\ee^{q \zeta + \xi_1^{-1} \zeta^3}E_1( q \zeta + \xi_1^{-1} \zeta^3) \,\dd\zeta + O(\ep^{4/3}), \label{eq:unif_tau}
\end{equation}
where $M = \delta/\ep^{2/3}$, $\psi = \arg \sqrt{\xi_1} = -\pi/4$, and $\zeta_0 = \sqrt{-q\xi_1}$. The two integrals make up the black contour consisting of straight lines at angle $\pi/4$ to the axes in figure~\ref{fig: Contours}, traversed in the sense indicated by the arrows. We now deform these contours around the branch cuts of $E_1$, the red curves in figure~\ref{fig: Contours}. The exponential integral has a discontinuity of $-2\pi\ci$ across these branch cuts. As a result the exponential integral in the integrand will contribute $2\pi\ci$ times the rest of the integrand when integrated along Hankel-like contours around the branch cuts. We deform twice, once using the two contours in (\ref{eq:unif_tau}), i.e.~the left and right wedge contours, and once deforming onto upper and lower wedge contours, then average. Since the integrand is $O(|\zeta|^{-2})$ for large $|\zeta|$, $M$ can be replaced by infinity. We obtain
\begin{equation}
\Ib \sim \ep^{2/3} 2\pi\ci \frac{\fb(\theta_a)}{\xi_1(\theta_a)} \left[\int_{-\zeta_0}^{-\ci\infty} - \int_0^{\infty\ee^{5\pi\ci/6}} - \int_{\zeta_0}^{\infty\ee^{\pi\ci/6}} \right]\zeta\ee^{q\zeta+\xi_1^{-1}\zeta^3} \,\dd\zeta \label{local contribution I1 po}
\end{equation} 
for $q > 0$ and
\begin{equation}
\Ib \sim\ep^{2/3} 2\pi\ci \frac{\fb(\theta_a)}{\xi_1(\theta_a)} \left[\int_{\zeta_0}^{-\ci\infty} - \int_0^{\infty\ee^{\pi\ci/6}} - \int_{-\zeta_0}^{\infty\ee^{5\pi\ci/6}} \right]\zeta\ee^{q\zeta+\xi_1^{-1}\zeta^3} \,\dd\zeta \label{local contribution I1 neg}
\end{equation}
for $q<0$.

The local contribution of $J$ comes from three intervals, which will be transformed into $\tau$- and then $\zeta$-integrals as above. We need to take into account
\begin{equation}
  -\lambda a d = -\sqrt{\xi_1}\sqrt{\tau} |q + \tau|, \label{eq:abds}
\end{equation}
in the $\tau$-integrals. The two cases shown in Figure~\ref{fig: Contours} correspond to $q > 0$, i.e.~$\theta_a > \theta_d$ (left panel), and $q < 0$, i.e.~$\theta_d > \theta_a$ (right panel). For $q > 0$ we take the intervals $(\theta_a,\theta_a+\delta)$, $(\pi +\theta_d, \pi + \theta_a)$ and $(\theta_d-\delta,\theta_d)$. These intervals map to $(-\infty,-q)$, $(0,\infty)$ and $(-q,0)$ in the $\theta$-plane. The $\sgn{d}$ term in (\ref{eq:J}) leads to a minus sign multiplying the first two contributions, while the absolute value in the exponent means that the first two terms have $-h(\tau)$ there, while the third has $h(\tau)$. In the $\zeta$-plane, we obtain the intervals $(0,\sqrt{\xi_1}\infty)$, $(\zeta_0,0)$ and $(\ci\sqrt{\xi_1}\infty,\zeta_0)$. Putting this all together and changing variable to obtain $h(\zeta)$ in all the exponentials, using the fact that the rest of the integrand is odd, leads to
\begin{equation}
  \Jb = \ep^{2/3} 2\pi\ci \frac{\fb(\theta_a)}{\xi_1(\theta_a)} \left[ \int_{0}^{-\infty\sqrt{\xi_1}} + \int_0^{-\zeta_0}  + \int_{\zeta_0}^{\infty\ci\sqrt{\xi_1}} \right] \zeta\ee^{(q\zeta+\xi_1^{-1}\zeta^3)} \,\dd\zeta + O(\ep^{4/3}).
\end{equation}
For $q < 0$ we take the intervals $(\theta_d,\theta_d+\delta)$,  $(\pi+\theta_a,\pi+\theta_d)$  and $(\theta_a-\delta,\theta_a)$. The transformed intervals are $(-q,\infty)$, $(0,-q)$ and $(-\infty,0)$, with $-h(\tau)$ in the first integrand and the third interval including a minus prefactor. The $\zeta$-intervals are then $(\zeta_0,\sqrt{\xi_1}\infty)$, $(0,\zeta_0)$ and $(\ci\sqrt{\xi_1}\infty,0)$. The same transformations give
\begin{equation}
  \Jb \sim \ep^{2/3} 2\pi\ci \frac{\fb(\theta_a)}{\xi_1(\theta_a)}\left[ \int_{-\zeta_0}^{-\infty\sqrt{\xi_1}} + \int_ 0^{\zeta_0}+ \int_{0}^{\infty\ci\sqrt{\xi_1}} \right] \zeta\ee^{(q\zeta+\xi_1^{-1}\zeta^3)} \,\dd\zeta.
\end{equation}

Combining the results for $\Ib$ and $\Jb$ leads
\begin{equation}
  \Gb \sim \ep^{2/3} 2\pi\ci \frac{\fb(\theta_a)}{\xi_1(\theta_a)} \int_{0}^{-\ci\infty} \zeta\ee^{(q\zeta+\xi_1^{-1}\zeta^3)} \,\dd\zeta = \fb(\theta_a) \hat G(q;\theta_a), \label{eq:onbeamas1}
\end{equation}
for all $q$, which defines a scalar function $\hat G$.. The contour of integration goes from the origin to $-\ci\infty$ and does not depend on $q$. The function $\hat G$ can be rewritten in a form closer to that in (\ref{eq:TSG}) from  \cite{Thomas:1972}:
\begin{equation}
	\hat G \sim \ep^{2/3} \frac{2\pi}{\abs{\xi_1(\theta_a)}^{1/3}} \int_{0}^{\infty} \zeta\ee^{-\zeta^3}\ee^{-\ci q \abs{\xi_1}^{1/3}\zeta} \,\dd\zeta =
	\pi (2\mu)^{2/3} \int_{0}^{\infty} \zeta\ee^{-\zeta^3}\ee^{-\ci p\zeta} \,\dd\zeta, \label{eq:onbeamas2}
\end{equation} 
with $p = q (2\omega/N)^{-1/3} (1 - \varpi^2)^{1/6}$. This integral is closely related to the Scorer function $\Hi(x)$,  a generalization of the Airy function. used by \cite{Voisin:2003}, Note that the real part of $\Gb$ is an even function of $q$, while the imaginary part is odd.

\begin{figure}
  \centering
  \includegraphics[width=\textwidth]{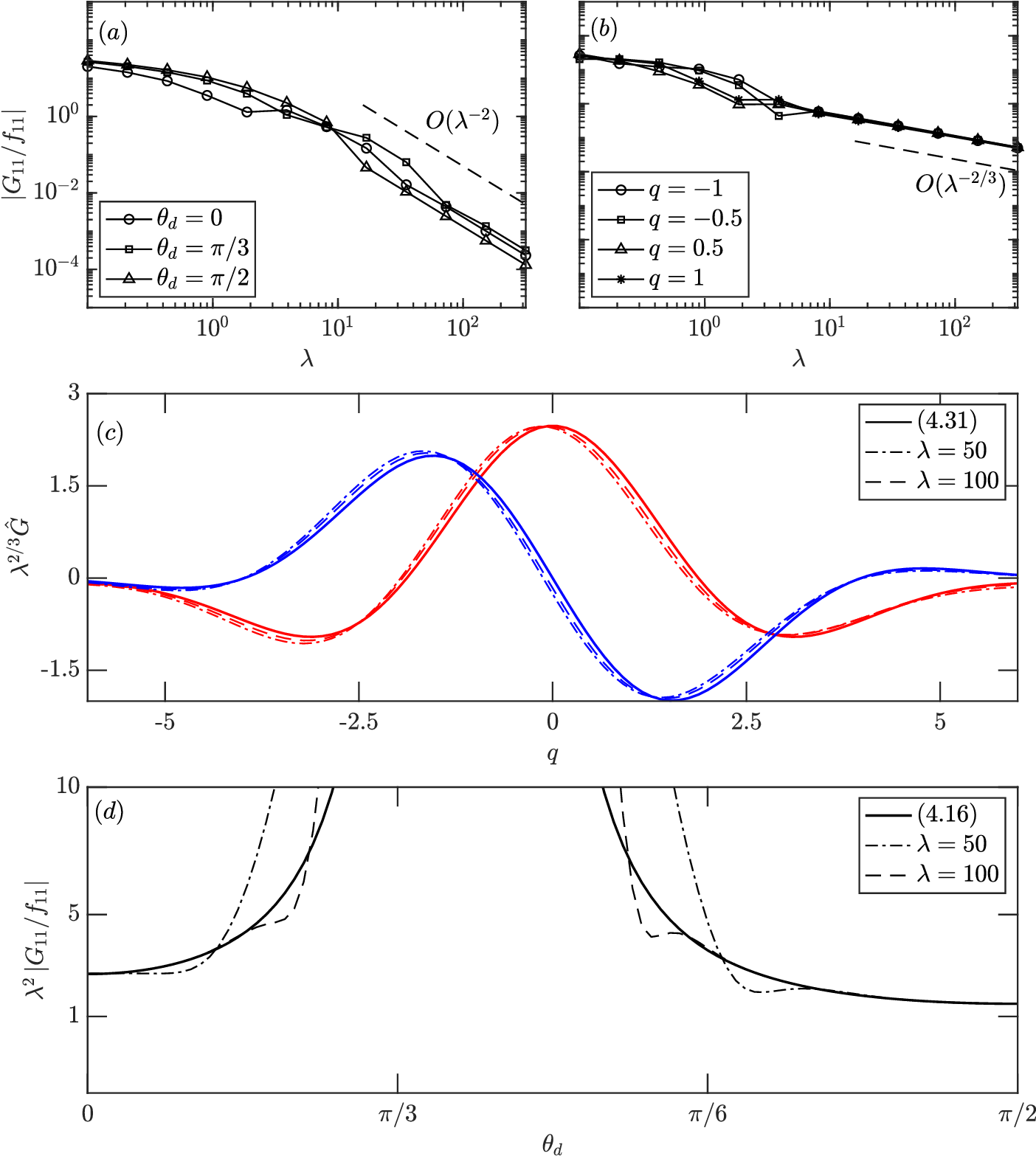}
  \caption{Top row: far-field behaviour as a function of $\lambda$ for $\omega/N = 0.8$; (a) off beam, (b) near beam. Middle row (c): rescaled behaviour of numerical and asymptotic solutions near the beam (red: real parts, blue: imaginary parts). Bottom row (d): off-beam.}
  \label{fig:IWas}
\end{figure}

Figure~\ref{fig:IWas} shows the asymptotic behaviour of $\Gb$ away
from and near to the beam, which agrees with the order of magnitudes
$\ep^2$ and $\ep^{2/3}$, along with a comparison of
(\ref{eq:onbeamas2}) with numerical calculations in the stretched
coordinate $q$ for various values of $\lambda$. The near-beam
asymptotic analysis of \cite{Thomas:1972}, using a boundary-layer
approach close to the internal wave angle $\theta_a$, does not provide
a full picture of the wave-field and is hard to extend to other
problems, which we do below.

A composite expansion can be obtained from combining (\ref{eq:offbeamas}) and (\ref{eq:onbeamas2}), subtracting off their overlap:
	\begin{align}
		\Gb \sim & \ep^{2/3} 2\pi \frac{\fb(\theta_a)}{\abs{\xi_1(\theta_a)}^{1/3}} \int_{0}^{\infty} \zeta\ee^{-\zeta^3}\ee^{-\ci q \abs{\xi_1}^{1/3}\zeta} \,\dd\zeta \nonumber \\
		& - \pi\ep^2\cot{\theta_a}\left[ \frac{\fb(\theta_a) }{\sin^2{(\theta_a - \theta_d)}} - \frac{\fb(\theta_a)}{(\theta_a - \theta_d)^2} + \frac{\fb(-\theta_a) }{ \sin^2{(\theta_a + \theta_d)}} \right]. \label{eq:compG}
	\end{align}
	This form suggests how to obtain a uniformly valid solution, which we now present.

\subsection{Uniform asymptotic expansion}
	\label{sec:unif}
The reason for the existence of two different asymptotic expansions so
far is the coalescence of singularities between $\Theta_a$ and $\Theta_d$. Hence, to find a uniform asymptotic expansion, we rescale the integration variable in the expansion \eqref{theta integral in I} about $\theta_a$ as $\theta = \theta_a + q \sigma \tau$, where $q = \sin{(\theta_a-\theta_d)}$ in the interval $(-\theta_a,\pi-\theta_a)$, and similarly about $\pi+\theta_a$ in the interval $(\pi-\theta_a,-\theta_a)$.

Following the approach in \S\,\ref{subsec:approach}, we decompose the integral into local and global contributions. The relevant expansion becomes
\begin{equation}
	\lambda ad \sim \lambda \sqrt{\xi_1(\theta_a) q\sigma \tau} [ q + q\sigma \tau\sqrt{1-q^2} ] = h(\tau), \label{lad tad eqn1}
\end{equation}
The leading local contribution of $\Ib$ from $\theta_a$ and $\pi +\theta_a$ is, using \eqref{lad tad eqn1},
\begin{equation}
	\Ib_{l,a}  =\;\sigma q\; {\fb(\theta_a)}\int_{-\delta/\sigma}^{\delta/\sigma} [\ee^{h(\tau)}E_1(h(\tau)) + \ee^{-h(\tau)}E_1(-h(\tau))] \,\dd\tau,
\end{equation}
where $\sigma$ and $\delta$ are chosen such that $\sigma \ll \delta$. Define $\zeta = \lambda \sqrt{\xi_1(\theta_a) \sigma\tau}$, so that
\begin{equation}
	\Ib_{l,a} =\ep^2\frac{2\fb(\theta_a)}{\xi_1(\theta_a)} q\left[\int_{{M} \ci\ee^{\ci \psi}}^{{M} \ee^{\ci \psi}} + \int_{-{M} \ci\ee^{\ci \psi}}^{-{M} \ee^{\ci \psi}}\right] \zeta\ee^{q^{3/2}(\zeta - \ep^2\phi(q)\zeta^3)}E_1(q^{3/2}(\zeta - \ep^2\phi(q)\zeta^3)) \,\dd\zeta,
\end{equation}
where $M = \sqrt{\abs{\xi_1(\theta_a)}}\; \delta^{1/2}/\ep$, $\phi(q) = -\sqrt{1-q^2}/\xi_1(\theta_a)$ and $\psi = \arg \sqrt{\xi_1(\theta_a)} = -\pi/4$. We can replace $M$ with $\infty$ in the above integrals, treating them as finite parts.
Following figure \ref{fig: Contours3} with appropriate deformation of
the contours of integration onto the branch cuts of $E_1$ (the red curves) results in
\begin{equation}
	\Ib_{l,a} =\ep^2 2\pi\ci \frac{\fb(\theta_a)}{\xi_1(\theta_a)} q\left[ -\int_{\zeta_0/\ep}^{\infty \ee^{\ci \psi_1}} + \int_{-\zeta_0/\ep}^{\infty \ee^{\ci\psi_2}} -  \int_0^{\infty \ee^{\ci\psi_3}} \right]\zeta\;\ee^{q^{3/2}(\zeta - \ep^2\phi(q)\zeta^3)} \,\dd\zeta, \label{uniform-Ia }
\end{equation}
where $\zeta_0 = (-\xi_1(\theta_a)/\sqrt{1-q^2})^{1/2}$, $\{\psi_1, \psi_2,\psi_3\} = \{\pi/6, 3\pi/2,5\pi/6\}$ when $q>0$ and $\{\psi_1, \psi_2\} = \{\pi/3, \pi,-\pi/3\}$ when $q<0$. 

\begin{figure}
	\centering
	\includegraphics[width=0.8\linewidth]{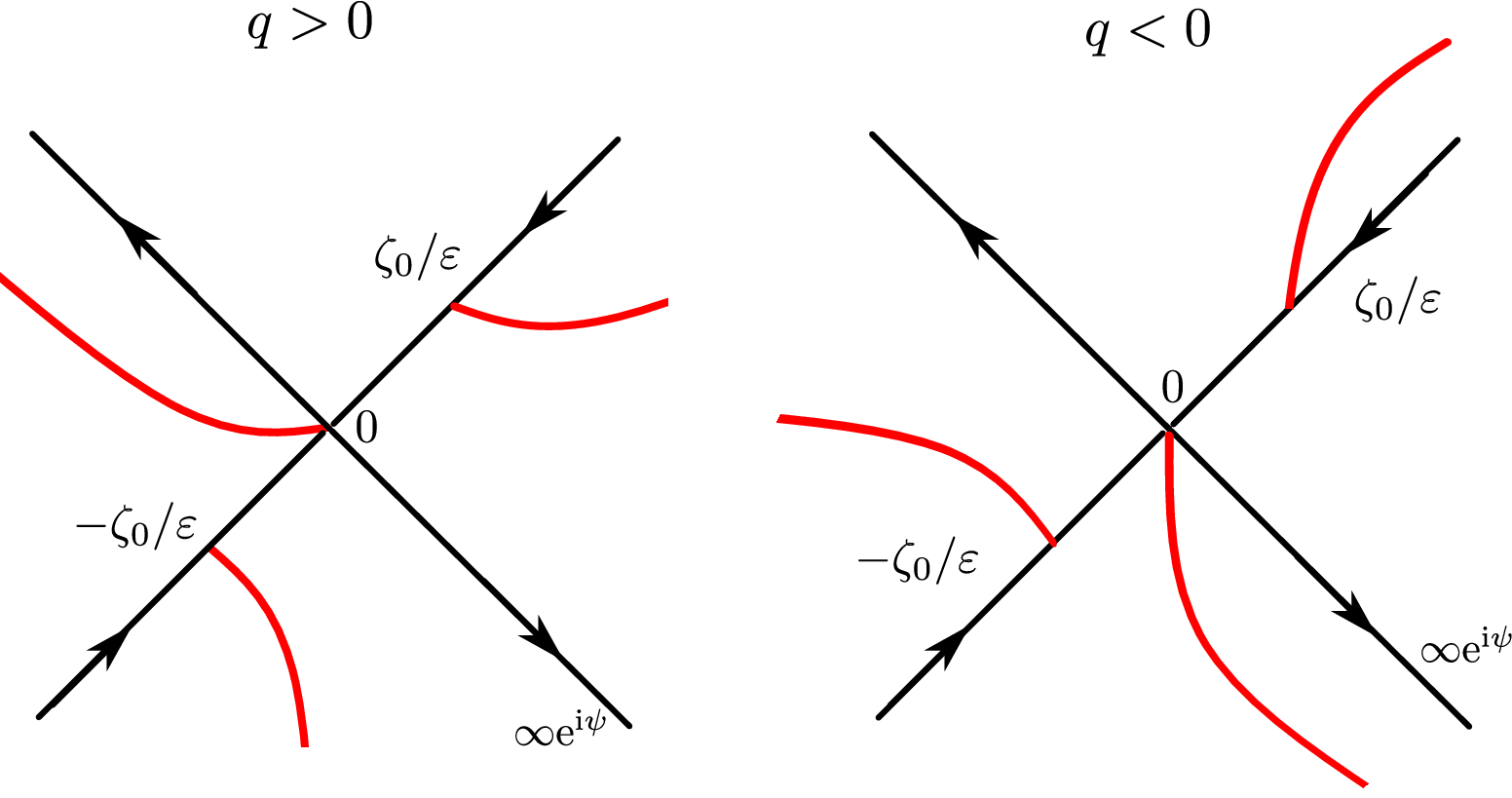}
	\caption{Schematic of integration contours (black lines) in the
    $\zeta$-plane and branch cut structure (red curves) of $\log{(q^{3/2}\zeta - \phi(q)\ep^2\zeta^3)}$ for $q \lessgtr 0$. The branch cuts extend to infinity along $\{\ee^{\ci \pi/6}, \ee^{\ci 5\pi/6}, \ee^{\ci 3\pi/2}\}$ for $q>0$ and $\{\ee^{\ci \pi/3}, \ee^{\ci \pi}, \ee^{-\ci\pi/3}\}$ for $q<0$.}
	\label{fig: Contours3}
\end{figure} 

The local contribution from $\Jb$ from $\{\theta_a,\pi+\theta_a\}$ is
\begin{equation}
	\Jb_{l,a} = \ep^{2} 2\pi\ci \frac{ \fb(\theta_a)}{\xi_1(\theta_a)} q\left[\int_{0}^{\infty\ee^{\ci5\pi/6}} + \int_{0}^{-\zeta_0/\ep} + \int_{\zeta_0/\ep}^{\infty \ee^{\ci\pi/6}}\right] \zeta\;\ee^{q^{3/2}(\zeta - \ep^2\phi(q)\zeta^3)} \,\dd\zeta,
\end{equation}
when $q>0$ and 
\begin{equation}
	\Jb_{l,a} = \ep^{2} 2\pi\ci \frac{ \fb(\theta_a)}{\xi_1(\theta_a)} q\left[\int_{0}^{\infty\ee^{-\ci\pi/3}} + \int_{0}^{-\zeta_0/\ep} + \int_{\zeta_0/\ep}^{\infty \ee^{\ci\pi/3}}\right] \zeta\;\ee^{q^{3/2}(\zeta - \ep^2\phi(q)\zeta^3)} \,\dd\zeta.
\end{equation}
when $q<0$. The total local contributions from $\{\theta_a,\pi+\theta_a\}$ become
\begin{equation}
	\Ib_{l,a} + \Jb_{l,a} = \ep^2 2\pi\ci \frac{\fb(\theta_a)}{\xi_1(\theta_a)}  \int_{0}^{-\ci\infty}\zeta\;\ee^{q\zeta-\ep^2\phi(q)\zeta^3} \,\dd\zeta,
\end{equation}
for all $q$.

Since there is no coalescence of singularities in this approach, the local contribution from $\{\theta_d,\pi+\theta_d\}$ can be computed using the results from the off-beam cases, leading to a contribution of $O(\ep^3)$. Hence the leading-order uniform expansion is
\begin{align}
	\Gb &= \ep^2 2\pi\ci \frac{\fb(\theta_a)}{\xi_1(\theta_a)}  \int_{0}^{-\ci\infty}\zeta\;\ee^{s_{\Delta}\zeta+\ep^2\xi^{-1}_1 c_{\Delta}\zeta^3} \,\dd\zeta -\ep^2 \pi \cot{\theta_a} \frac{\fb(-\theta_a)}{\sin^2{(\theta_a + \theta_d)}} + O(\ep^{4/3}, \ep^3) \label{sec:unifexp}, \\
	&= \mu^2 \pi\; \fb(\theta_a)
   \int_{0}^{\infty}\zeta\;\ee^{-\ci s_{\Delta}\zeta+(\mu^2/2)
   c_{\Delta}\zeta^3} \,\dd\zeta -\mu^2 \pi
   \frac{\fb(-\theta_a)}{\sin^2{(\theta_a + \theta_d)}} + O(\mu^{4/3},
   \mu^3) \label{sec:unifexp2}
\end{align}
with $s_\Delta = \sin{(\theta_a-\theta_d)}$ and $c_\Delta = \cos{(\theta_a-\theta_d)}$.

\section{Special cases}
\label{sec: special cases}

\subsection{Steady case: $\omega \to 0$}
\label{sec:steady}
This corresponds to $\theta_a \to \pi/2$. Off the beam, expanding (\ref{eq:GBasorigv}) in this limit gives
\begin{equation}
  \Gb \sim -2\mu^2 \left(
    \begin{array}{rr}
      1/\hat{z}^2 & 2 \varpi^2 \hat{x}/\hat{z}^3 \\
      2 \varpi^2 \hat{x}/\hat{z}^3 &  \varpi^2 /\hat{z}^2
    \end{array} \right).
\end{equation}
We have kept the dominant term for each matrix element. The dominant response is $G_{11}$, corresponding to horizontal response due to horizontal forcing. This expansion is not uniformly valid close to the horizontal axis, corresponding to the beam. The near-beam result (\ref{eq:onbeamas2}) and the uniform result (\ref{sec:unifexp2}) can be used, employing the variables $\mu$ and $p$.

\subsection{Critical case: $\omega \to N^-$}
\label{sec:critical}
This corresponds to $\theta_a \to 0$. Off the beam (\ref{eq:GBasorigv}) then becomes
\begin{equation}
  \Gb \sim -2 \mu^2 \left(
  \begin{array}{rr}
  	(1 - \varpi^2)/\hat{x}^2 & 2(1 - \varpi^2) \hat{z}/\hat{x}^3 \\
  2(1 - \varpi^2) \hat{z}/\hat{x}^3	& 1/\hat{x}^2
  \end{array} \right), \label{eq:critG}
\end{equation}
when $\mu \ll 1$, again keeping the dominant term for each matrix element. The dominant response is $G_{22}$, corresponding to vertical response due to vertical forcing.  This expansion is not uniformly valid close to the vertical axis, corresponding to the beam. The near-beam result (\ref{eq:onbeamas2}) and the uniform result (\ref{sec:unifexp2}) can be used, employing the variables $\mu$ and $p$.

\subsection{Evanescent waves: $\omega > N$}
The difference from the results of \S\,\ref{sec: 2D Gfn} is that the function $a(\theta)$ no longer has real zeros and the $\theta_a$ are off the real axis. This leads to
\begin{equation}
  J = \begin{cases}
    0 & \mbox{if $d<0$}, \\
    \ci\pi\ee^{-\lambda a d} & \mbox{if $d>0$}.
  \end{cases}
\end{equation}
The procedure for numerical computations is the same as before, except
that if logarithmic singularities are removed from the integrand there
are only two, at $\theta_d$ and $\pi + \theta_d$.

\begin{figure}
	\centering
	\includegraphics[width=\textwidth]{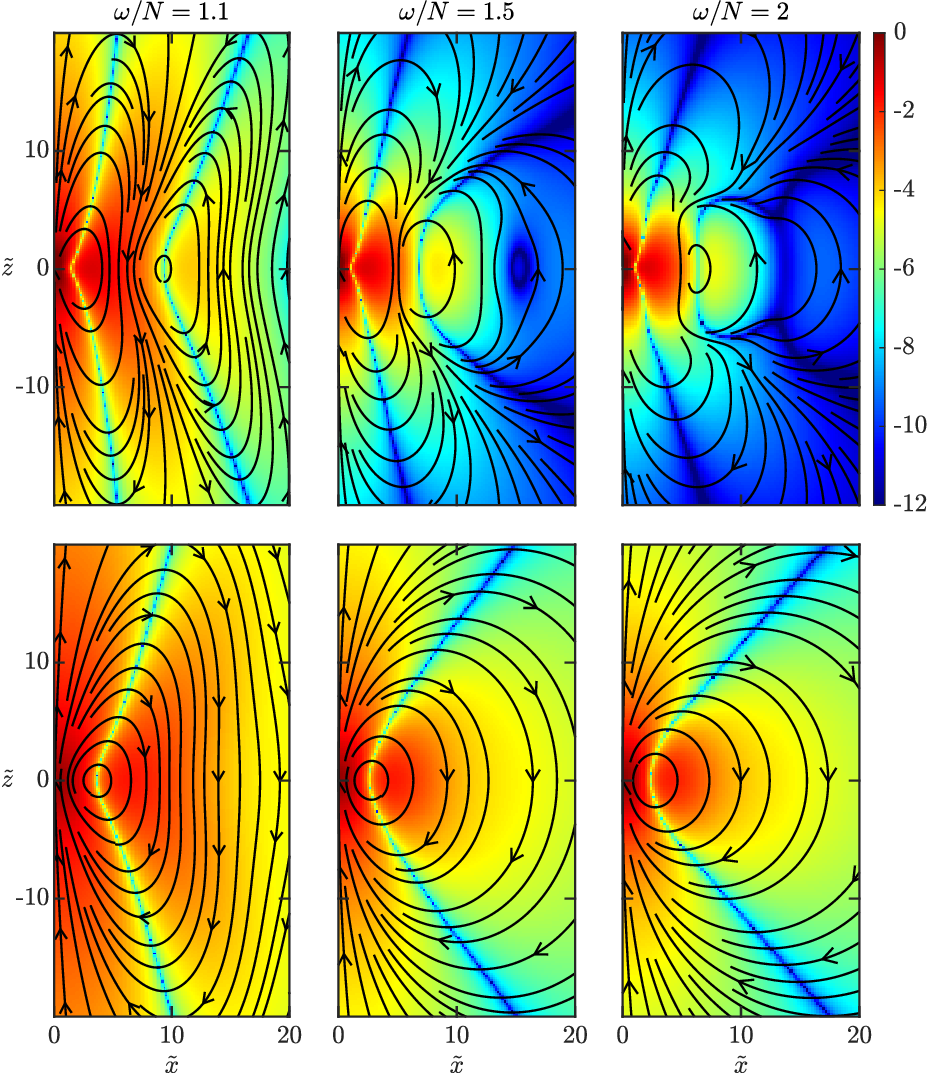}
	\caption{Evanescent response for unit positive vertical forcing. Top row: real part, bottom row:
          imaginary part. Left to right: $\omega/N = 1.1$, $1.5$, $2$. Colour: $\log{\abs{u_3}}$. Black lines: streamlines with the flow direction indicated by arrows.}
	\label{fig:2D EvanWaves}
\end{figure}

Figure~\ref{fig:2D EvanWaves} shows the logarithmic vertical velocity
$\log{\abs{u_3}}$ generated by a unit vertical forcing at the origin
for $\omega/N = 1.1$,$1.5$, $2$  as a colour plot, along with
streamlines. Unlike the internal wave case, we observe alternating
near-vertical cells in the real part and a single cell in the
imaginary part. Numerically we find that  for large $\lambda$, $\Imag\;\Gb\sim O(\lambda^{-2})$ and $\Real\;\Gb \sim O(\lambda^{-4})$. 

The asymptotic analysis follows the off-beam case of internal wave as above. The range of integration is divided into regions around and far from the zeros of $d(\theta)$ at $\Theta_d$. These are the only real zeros and there is hence no coalescence of singularities unlike internal waves, so there is no equivalent to on-beam and off-beam regions. Since the zeros of $a(\theta)$ are off the real axis, they do not appear in the local contributions, while the total local contributions from zeros of $d(\theta)$ to $\Ib$ and $\Jb$ are $O(\ep^3)$. The only non-zero contribution at $O(\ep^2)$ comes from the global contribution of $I$, which is no longer zero, but becomes, from (\ref{Global contribution series}),
\begin{align}
		\Gb = & -\ep^2 \oint \frac{\fb(\theta)}{a^2(\theta) d^2(\theta)} \,\dd\theta + O(\ep^3), \\
		\sim & \ep^2 \pi\cot{\theta_a}\left[\frac{\fb(\theta_a)}{\sin^2{\left(\theta_d-{\theta_a}\right)}} + \frac{\fb(-{\theta_a})}{\sin^2{\left(\theta_d+{\theta_a}\right)}}\right], \label{Evan asym result}
	\end{align}
where now $\theta_a = \ci\cosh{(\omega/N)}$ is purely imaginary. The off-beam asymptotic approximation of internal waves in (\ref{eq:offbeamas}) is similar to (\ref{Evan asym result}), up to a sign, and they result from local and global contributions respectively. At leading order, the off-beam wave field is purely real for $\omega<N$, while the evanescent wave field is purely imaginary. This is expected because, as $\omega$ approaches $N$ from below, the singularities of $\Theta_a$ coalesce at $\theta_a = 0$ and $\pi$ (see the right-hand panel of Figure~\ref{fig: theta angles}), so that the corresponding branch points of $E_1$ coincide. When $\omega$ increases beyond $N$, the branch points become conjugate pairs and move away from real axis. This could be analyzed using analytical continuation  in the complex variable $\omega/N$ of the  viscous internal wave Green's function, which is beyond the scope of this paper.  

\begin{figure}
  \centering
  \includegraphics[width=\textwidth]{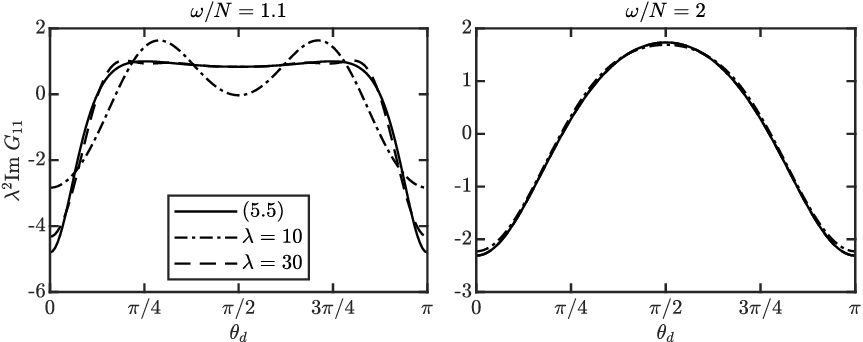}
  \caption{Rescaled  Green's function component $\lambda^2 \Imag{\; G_{11}}$ for  $\omega/N=1.1$, $2$ and$\lambda = 10$, $30$, as a function of $\theta_d$. Solid lines: asymptotic approximation from \eqref{Evan asym result}; broken lines: numerical results.}
  \label{fig:2D EvanWaves asym}
\end{figure}

Figure~\ref{fig:2D EvanWaves asym} compares the leading-order result from (\ref{Evan asym result}) to the numerical calculations for $\omega/N = 1.1$ and $2$ with $\lambda = 10$ and $30$. The asymptotic approximation improves as $\omega$ increases.
 
\section{Green's function for finite Pr}
\label{sec: 2D Gfn Pr}
\subsection{Integral representation}
\begin{figure}
\begin{center}
  \includegraphics[width=\textwidth]{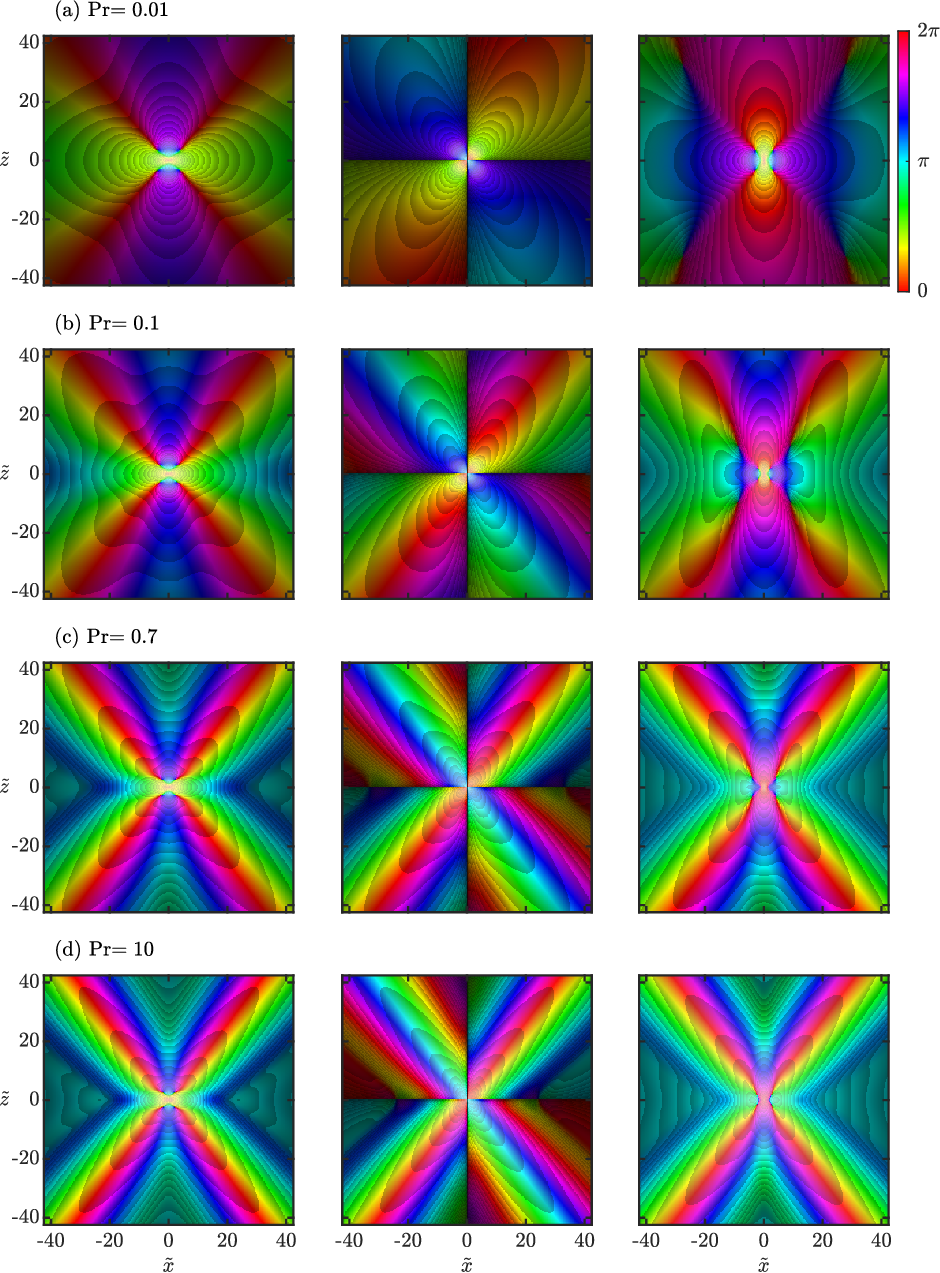}
        \caption{Phase portrait of $G_{ij}(\xb,\xb_0)$ for $\omega = 0.8$, $N=1$ and $\Pra  = 0.01$, $0.1$, $0.7$, $10$ (colour bar: phase angle).}
        \label{fig:2D Greens function Pr}
        \end{center}
    \end{figure}

With non-zero density diffusion, the integral representation of the Green's function takes the form
\begin{equation}
   \Gb  = \oint \fb(\theta) \,\dd\theta \int_0^\infty  \frac{\gamma \kappa }{(\kappa^2 -\ci)\gamma +\ci\left({N^2}/{\omega^2}\right) c^2}\ee^{\ci\lambda\kappa(c\tilde{x} + s\tilde{z})} \,\dd\kappa. 
\end{equation}
Using partial fractions, the $\kappa$-integral takes the form
\begin{equation}
    K^D = \int_0^\infty  \left[\frac{h_1(\theta;\Pra)}{\kappa^2 + a^2_1(\theta;\Pra)} + \frac{h_2(\theta;\Pra)}{\kappa^2 + a_2^2(\theta;\Pra)}\right] \kappa \ee^{\ci\lambda d(\theta)\kappa} \,\dd\kappa,
\end{equation}
where the quantities $a_{1,2}(\theta;\Pra)$ and $h_{1,2}(\theta;\Pra)$ are given by
\begin{align}
    a^2_{1,2}(\theta;\Pra) &= -\frac{\ci}{2}\left[(1+\Pra) \mp \sqrt{ (1-\Pra)^2 + 4\Pra \frac{N^2}{\omega^2}c^2} \right], \label{finite Pr a1, a2} \\
        h_1(\theta;\Pra) &= \frac{a^2_1 + \ci \Pra}{a^2_1 -a^2_2}, \qquad  h_2(\theta;\Pra) = \frac{a^2_2 + \ci \Pra}{a^2_2 -a^2_1}, \label{finite Pr h1, h2}
\end{align}
with the upper sign corresponding to $a_1$ and the lower sign to $a_2$.
Note that $h_1(\theta;\Pra)$ and $h_2(\theta;\Pra)$ are finite,
$a_1(\theta;\Pra)$ vanishes for $\theta \in \Theta_a$ given in
(\ref{a0, b0 eqns}), and $a_2(\theta;\Pra) \neq 0$. We can construct the solution using (\ref{kappa integral IW}) as
\begin{equation}
    h_1K(a_1,d) + h_2 K(a_2,d), \label{eq:KPra}
\end{equation}
where the terms $K(a_1,d)$ and $K(a_2,d)$ behave similarly to internal wave and to unsteady Stokes flow  respectively, as can be seen by examining the limits $\Pra \to 0$ ($h_1\to 0$) and $\Pra \to \infty$ ($h_2\to 0$).  We can hence split the Green's function into two terms as
\begin{equation}
  \Gb = h_1 \Gb_1 + h_2 \Gb_2 = \Gb^W + \Gb^S,
\end{equation}
corresponding to wave and Stokes terms given by the integrals of the two terms in (\ref{eq:KPra}). The Green's function is shown for different Prandtl numbers in figure~\ref{fig:2D Greens function Pr} using the same numerical approach as before.

\subsection{Asymptotic analysis for finite $\Pra$}

For the off-beam asymptotic behaviour, one can follow the same analysis as in \S\,\ref{subsec: local contribution off}. We find that $\Gb^S = O(\ep^{3})$ since $a_2(\theta)$ has no zeros. The leading-order behaviour comes at $O(\ep^2)$ from $\Gb^W$. Near the point $\theta_a$, in (\ref{lam ad original}) $\xi_1$ is replaced by $\xi_1/(1+\Pra^{-1})$, while $h_1(\theta_a;\Pra) = (1 + \Pra^{-1})^{-1}$. The Prandtl-dependent terms then cancel and one recovers (\ref{eq:offbeamas}). This result is independent of $\Pra$. Similarly the evanescent case recovers (\ref{Evan asym result}). 
  
For the on-beam asymptotics, excluding the case $\Pra \ll 1$, the analysis follows that of \S\,\ref{subsec: local contribution} for the wave-like behaviour coming from the $h_1$ term. The only difference is the new term in $\xi_1$, so that the result is
\begin{equation}
    \Gb^W \sim \ep^{2/3} 2\pi\frac{\fb(\theta_a)}{\abs{\xi_1(\theta_a)}^{1/3}}(1 + \Pra^{-1})^{-2/3} \int_0^{\infty}\zeta \ee^{-\zeta^3} \ee^{-\ci q\abs{\xi_1}^{1/3}(1 + \Pra^{-1})^{-1/3} \zeta} \,\dd\zeta.
\end{equation}
The prefactor $(1+\Pra^{-1})^{-2/3}$ decreases as the Prandtl number decreases, with the results that the amplitude decreases and the beam width increases. The uniform result (\ref{sec:unifexp}) is modified in the same way.

\section{Anisotropic Brinkman flow}
\label{sec: brinkman}
\subsection{Formulation}

The dimensionless governing equations for anisotropic Brinkman flow are \cite[]{Kohr:2007}
\begin{equation}
    -\nabla q + (\nabla^2 - \mathbf{K})\ub  + \fb= \mathbf{0}, \qquad \nabla\cdot \ub = 0, \label{Brinkman eqns 1}
\end{equation}
where $q$ is pressure, $\fb = \gb \delta(\xb -\xb_0)$ and $\mathbf{K}$ is the non-dimensional diagonal permeability tensor of the porous medium. The permeability tensor takes the form
\begin{equation}
    \mathbf{K} = \left(\begin{array}{ll}
        \chi_1 & 0 \\
        0 & \chi_3
    \end{array}\right) \equiv -\frac{\ci \omega}{\nu}\left(\begin{array}{ll}
        1 & \quad 0 \\
        0 & 1 -  \left({N}/{\omega}\right)^2
    \end{array}\right).
\end{equation}
where the second expression corresponds to the viscous internal
wave case with $\Pra \to \infty$ discussed previously. For Brinkman
flow $\chi_1$ and $\chi_3$ are positive real values, while for viscous
internal waves, the corresponding matrix entries are purely imaginary,
with large magnitudes for small viscosity. For Brinkman flow all
variables are real from \eqref{Brinkman eqns 1}. Without loss of
generality we take $\chi_3>\chi_1$.

We again write 
\begin{equation}
u_i = \frac{1}{4\pi}G_{ij}(\xb,\xb_0) g_j.
\end{equation}
Then the Fourier transforms of the anisotropic Brinkman's Green's function's components are
\begin{align}
  \mathcal{F} G_{1j} =  4\pi \left(\delta_{1j} - \frac{k_1k_j}{\abs{\kb}^2}\right) \frac{1}{\abs{\kb}^2 + \chi_1 + (\chi_3-\chi_1) {k_1^2}/{\abs{\kb}^2}}\ee^{-\ci\kb\cdot\xb_0}, \qquad j = 1,2, \\
  \mathcal{F} G_{2j} =  4\pi \left(\delta_{2j} - \frac{k_2k_j}{\abs{\kb}^2}\right) \frac{1}{\abs{\kb}^2 + \chi_3 + (\chi_1-\chi_3) {k_2^2}/{\abs{\kb}^2}}\ee^{-\ci\kb\cdot\xb_0}, \qquad j = 1,2.
\end{align}
Taking the inverse Fourier transform and using polar coordinates $(k_1,k_2) = \kappa(\cos{\theta},\sin{\theta}) \equiv \kappa(c,s)$, with $a^2 = \chi_1 s^2 + \chi_3 c^2$, $d = (\hat{x} \cos{\theta} + \hat{z}\sin{\theta})\equiv  \sin{(\theta - {\theta}_d)}$, $\lambda = |\xb - \xb_0|$ and $\hat{\xb} = (\xb-\xb_0)/\abs{\xb-\xb_0}$, gives
	\begin{equation}
		\Gb = \oint  \fb(\theta)  K^B(a(\theta),d(\theta)) \,\dd\theta. \label{Brink Gij int eqn}
	\end{equation}
	The  $\kappa$-integral for anisotropic Brinkman flow is
	\begin{equation}
		K^B(a(\theta),d(\theta)) = \int_0^\infty \frac{ \kappa }{\kappa^2 + a^2(\theta)} \ee^{\ci \lambda d(\theta)\kappa } \,\dd\kappa.
	\end{equation}
	Without loss of generality we take $a  = \sqrt{\chi_1 s^2 + \chi_3 c^2} > 0$. The complex zeros of $a(\theta)$ are situated at
	\begin{equation}
		\Theta_{a} = \left\{\frac{\pi}{2}\pm \ci \tanh^{-1}\left(\frac{\chi_1}{\chi_3}\right)^{1/2}, -\frac{\pi}{2} \pm \ci \tanh^{-1}\left(\frac{\chi_1}{\chi_3}\right)^{1/2}\right\},
	\end{equation}
while $d(\theta)$ has real zeros at $\theta=\theta_d$, $\pi + \theta_d$. Since the velocity fields in \eqref{Brinkman eqns 1} are real for arbitrary real forcing $\fb$, $K^B$ in (\ref{eqA:BKB}) can be replaced by
\begin{equation}
  K^B({a}, d) = \frac{1}{2} \Real{[\ee^{ad}\tilde E_1(ad) +  \ee^{-ad} \tilde E_1(-ad)]} , \label{K^B eqn brink}
\end{equation}
with the exponential term from the residue dropped since it is purely imaginary and hence vanishes on integration.
  
\subsection{Numerical results}
The same  numerical approach as for evanescent internal waves can be applied. Figure~\ref{fig: Brinkman 2D 2,3 Greensfunction} shows the logarithmic vertical velocity, $\log{|u_3|}$, generated by a unit vertical forcing at the origin for $\chi_1 = 1$, varying $\chi_3,$ as a colour plot, along with streamlines. As $\chi_3$ increases, so that the system becomes more anisotropic, the flow becomes weaker away from the origin in $z$.  This is because $u_3$ must decrease  as $\chi_3$ increases to balance the pressure gradient. The resulting circulation cells are also more limited in $z$.

\begin{figure}
  \centerline{\includegraphics[width=\textwidth]{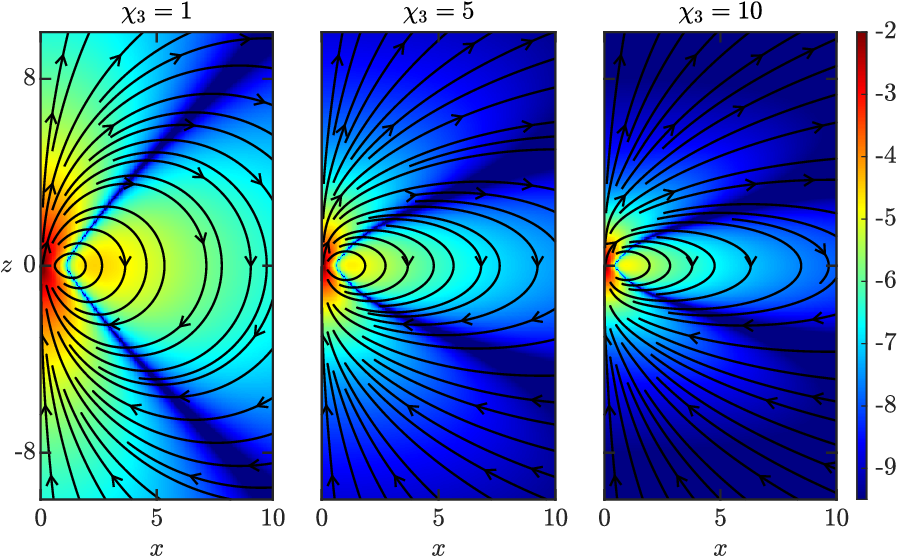}}
        \caption{Anisotropic Brinkman solution for unit positive vertical forcing. Colour: $\log{|u_3|}$ with $\chi_1 = 1$ and $\chi_3 = 1$, $5$ and $10$. Black lines: streamlines with the flow direction indicated by arrows.}
        \label{fig: Brinkman 2D 2,3 Greensfunction}
\end{figure}

\subsection{Asymptotic analysis}
\label{sec: 2D brink asym}

The asymptotic analysis for large $\lambda$ of anisotropic Brinkman flow is almost the same as for evanescent waves. The global contribution to $\Gb^B_g$ once again can be obtained from the expansion (\ref{Global contribution series}). We compute the integral using residue calculus. The poles at $\theta_d$ do not contribute, as for \eqref{subsec: global contribution}. There are now non-zero contributions from the simple poles at $\Theta_a$, giving
\begin{equation}
I^B_g \sim  4\pi \Imag\;{\left[\sum_{q=1}^\infty \ep^{2q} \frac{\Gamma(2q)}{\Gamma(q)} \frac{\dd^{(2q-1)}}{\dd\theta^{2(q-1)}} \left(\frac{(\theta - \theta_a)^q \fb(\theta)}{a^{2q}(\theta) d^{2q}(\theta)}\right)_{\theta = \theta_a}\right]},
\end{equation}
where $\theta_a$ is one of the zeros of $a(\theta)$ in the upper half-plane. The leading term is
\begin{equation}
\Ib^B_g \sim  \ep^2\frac{4\pi}{(\chi_1-\chi_3)}  \frac{\fb(\theta_a)}{\sin{2\theta_a} \sin^2{\theta_a}}. \label{eq:Brinkleadasymp}
\end{equation}

The local contribution now arises only from the vicinity of $\theta_d$ and $\pi + \theta_d$. The same expansion (\ref{lad theta_d exp}) as before holds, while (\ref{eq:Ibld}) is changed to
\begin{equation}
    \Gb^B_l = \ep {\fb(\theta_d)}\int_{-\delta/\ep}^{\delta/\ep} [\ee^{h(\tau)} \tilde E_1(h(\tau)) + \ee^{-h(\tau)} \tilde E_1(-h(\tau))] \,\dd\tau + \cdots,    	
  \end{equation}
Because of the properties of $\tilde E_1$, this integral can be split into two parts, one with a branch cut below the real axis and one with a branch cut above. The former part integrates to zero when closed in the lower half-plane, while the latter part integrates to zero when closed in the upper half-plane. In fact, all subsequent terms in the expansion are given by the same terms in the integrand multiplied by $\tau^n$, and vanish as finite-part integrals in the same way. Hence $\Gb^B_l = O(\ep^\infty)$.

\begin{figure}
	\begin{center}
		\centerline{\includegraphics[width=4in]{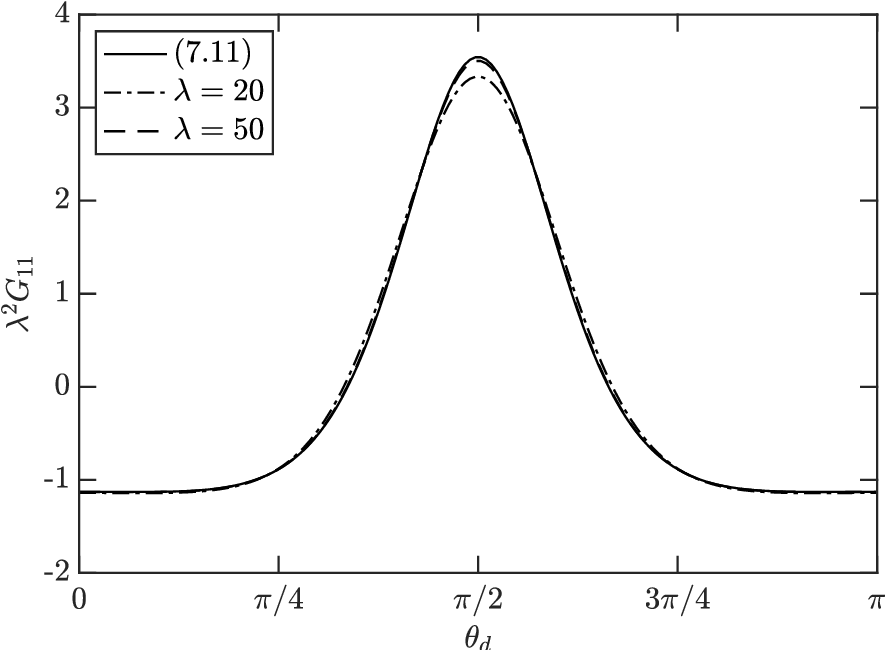}}
		\caption{Rescaled  Green's function component $\lambda^2 G_{11}$ for $\chi_1 = 1$, $\chi_3 = 5$, $\lambda = 20$, $50$, as a function of $\theta_d$. Solid line: exact result, broken lines: asymptotic approximation.}
		\label{fig: brinkman asym}
	\end{center}
\end{figure}

The leading-order asymptotic result (\ref{eq:Brinkleadasymp}) is compared to numerical results in Figure~\ref{fig: brinkman asym}. The solution is largest in the horizontal direction $(\theta_d = \pi/2)$.

\section{Conclusion}
\label{sec:conc}

We have derived a new formulation for the Green's function for viscous
internal waves, which allows efficient numerical computation and
rigorous asymptotic analysis. It is suitable for use in BIM
calculations, which might for example include looking at the ``critical slope singularity'' situation examined by \cite{Ledizes2024}. The two-dimensional solution is given by an angular integral over the unit circle with logarithmic singularities. Numerical calculations are straightforward, and can be made very accurate by removing singularities from the integrand.

The asymptotic behaviour of the resulting Green's function components
can be obtained using ``divide and conquer''methods,  with the use of the Hadamard finite part integral simplifying the analysis. Unlike previous work, we do not approximate the solution to compute the asymptotic behaviour, allowing us to obtain a uniform asymptotic expansion.

The resulting wave field is concentrated along beams at the internal
wave angle, along which the solution decays more slowly than in the
rest of the the domain. Previous results are recovered in the vicinity
of the beam, characterized by a decay rate of $\lambda^{-2/3}$. When
the internal frequency is larger than the buoyancy frequency, an
evanescent solution is obtained, whose far field was also
analysed. The special cases of zero frequency and critical frequency
($\omega = N$) can be obtained from the general result, showing the
utility of the variable $\mu$.

Density diffusion was also taken into account. The wave field now can be viewed as a superposition of wave and Stokes terms. The asymptotic behaviour is dominated by the wave term. The effect of finite Prandtl number is to decrease the overall amplitude of the beam by a factor of $(1+\Pra^{-1})^{-2/3}$ and to increase the beam width by a factor of $(1+\Pra^{-1})^{1/3}$. Off the beam and in the evanescent case, density diffusion does not appear in the leading-order asymptotic approximations. For $\Pra \gtrsim O(1)$, the leading order off-beam behaviour is purely real for internal waves and purely imaginary for evanescent solutions.

We also computed the Green's function for anisotropic Brinkman flow using the same approach. A full asymptotic series can be obtained in the far-field limit. The asymptotic behaviour is similar to the evanescent case. Both fields are dominated by a circulation cell.

This approach can be extended to the three-dimensional case, which also allows rotation to be included. This will be the subject of future work.  Asymptotic calculations then lead to integrals over a unit sphere, with singular curves on the surface of the sphere rather than singular points.

%
\begin{bmhead}[Acknowledgements.]
  The authors are grateful to Bruno Voisin for helpful discussions on this topic and for pointing them to hard-to-find papers.
\end{bmhead}
\begin{bmhead}[Funding.]
  This research was partly funded by a NISEC award from the Office of
  Naval Research through the University of California, San Diego.
\end{bmhead}
\begin{bmhead}[Declaration of interests.]
  The authors report no conflict of interest.
\end{bmhead}
%
%
\begin{bmhead}[Author ORCIDs.]
  Saikumar Bheemarasetty, https://orcid.org/0009-0004-4411-1818; Stefan G. Llewellyn Smith, https://orcid.org/0000-0002-1419-6505
\end{bmhead}
%


\begin{appen}


\section{$\kappa$-integrals}
\label{appendix: 2D IW kappa integrals}
\subsection{Internal wave case ($a^2$ imaginary)}
The $\kappa$-integral (\ref{kappa integral IW}) is
\begin{align}
   K &= \frac{1}{2} \int_0^{\ci d\infty}\frac{\ee^\xi}{\xi-ad} \,\dd\xi + \frac{1}{2} \int_0^{\ci d\infty}\frac{\ee^\xi}{\xi+ad} \,\dd\xi \\
    &=  \frac{1}{2} \ee^{ad}\int_{ad}^{ad-\ci d\infty} \frac{\ee^{-\eta}}{\eta} \,\dd\eta + \frac{1}{2} \ee^{-ad}\int_{-ad}^{-ad-\ci d\infty} \frac{\ee^{-\eta}}{\eta} \,\dd\eta,  \label{I with vertical contours}
\end{align}
with $\ci d \kappa = \xi$. The integrand $\eta^{-1} \ee^{-\eta}$ has a singularity at $\eta = 0$ and the integration contours are straight vertical lines, going vertically down if $d>0$ and vertically up if $d<0$.

For $d>0$ and $\arg a = \pi/4$, the lower limits of integration $\pm ad$ lie in the first and third quadrants, respectively, and the contours of integration can be rotated anti-clockwise by an angle of $\pi/2$ to be horizontal. Then the integrals become exponential integrals, so that
\begin{equation}
    K = \frac{1}{2} [\ee^{ad}E_1(ad) +  \ee^{-ad}E_1(-ad)] = I, \label{eqA:Idef}
\end{equation}
which defines the function $I$.

For $d>0$ and $\arg a = -\pi/4$, the lower limits of integration $\pm ad$ lie in the fourth and second quadrants, respectively, and as the contours of integration are rotated through $\pi/2$, one obtains exponential integrals along with an additional contribution to the second integral from the residue at the origin. Hence
\begin{equation}
    K =\frac{1}{2} \ee^{ad}E_1(ad) +  \frac{1}{2}\ee^{-ad}[E_1(-ad) + 2\pi\ci] = I + \ci\pi\ee^{-ad}.
\end{equation}

For $d<0$, the contours need to be rotated by $\pi/2$ clockwise, and the residue contribution comes from the first integral when $\arg a = \pi/4$. Then
\begin{equation}
    K = I - \ci\pi\ee^{ad}
\end{equation}
for $\arg a = \pi/4$, and $K = I$ for $\arg a = -\pi/4$.
More compactly,
\begin{equation}
    K = I + \begin{cases} 0 &\mbox{if $\sgn{d}\arg a \in (0,\pi/2)$}\\
\sgn{d}(\ci\pi)\ee^{-a|d|} & \mbox{if $\sgn{d}\arg a \in (-\pi/2,0)$}. 
\end{cases}
\end{equation}

\subsection{Brinkman flow ($a > 0$)}
\label{appendix: kappa int Brinkman}
In anisotropic Brinkman flows, the same contour deformation approach gives
\begin{equation}
  K^B = I^B  + \sgn{d}(\ci\pi) \ee^{-a|d|}, \label{eqA:BKB}
\end{equation}
where $I^B$ is the integral defined in (\ref{eqA:Idef}) with the function $E_1(z)$ replaced by
\begin{equation}
  \tilde E_1(z) = \frac{1}{2} E_1(z + \ci 0) + \frac{1}{2} E_1(z - \ci 0).
\end{equation}
For negative real $x$, $\tilde E_1(x) = -\Ei{(-x)}$ while for any
other argument, $\tilde E_1(x) = E_1(x)$.
\end{appen}

\bibliographystyle{jfm}
\bibliography{IGW_viscous}

\end{document}